\documentclass[aps,prb,twocolumn,superscriptaddress,longbibliography]{revtex4-2}
\usepackage[colorlinks=true,linkcolor=blue,citecolor=blue]{hyperref}
\usepackage{amsmath,amssymb}
\usepackage[utf8]{inputenc}
\usepackage[T1]{fontenc}
\usepackage{url}
\usepackage{adjustbox}
\usepackage{subcaption}
\usepackage{graphicx}
\usepackage{bm}
\usepackage{bbm}
\usepackage{xcolor}
\usepackage{color}
\usepackage{ulem}
\usepackage{subfloat}
\usepackage{ragged2e}
\usepackage{caption}
\usepackage{braket}
\DeclareCaptionFormat{myformat}{\justifying#1#2#3}
\newsavebox{\measurebox}
\begin{document}

\title{Superconducting $p$-wave pairing effects on one-dimensional non-Hermitian quasicrystals with power law hopping}%

\author{Shaina Gandhi}
\email{p20200058@pilani.bits-pilani.ac.in}
\affiliation{Department of Physics, Birla Institute of Technology and Science, Pilani 333031, India} 

\author{Jayendra N. Bandyopadhyay}
\email{jnbandyo@gmail.com}
\affiliation{Department of Physics, Birla Institute of Technology and Science, Pilani 333031, India}

\begin{abstract}
We study the effects of superconducting $p$-wave pairing on the non-Hermitian Aubry-Andr\'e-Harper model with power-law hopping. For the case of short-range hopping, weak pairing leads to oscillating quasi-Majorana zero modes, turning to edge-localized Majorana zero modes as pairing strength increases. For the case of long-range hopping, we observe the emergence of massive Dirac modes having oscillatory behavior, similar to Majorana modes with weak pairing. The massive Dirac modes localize at the edges as the pairing strength grows. The superconducting pairing spoils the plateaus observed in the fractal dimension of all the energy eigenstates of the Aubry-Andr\'e-Harper model with power-law hopping. The number of plateaus decreases with the increasing pairing strength for the weak non-Hermiticity in the system. The phase diagram of the system reveals that real and complex energy spectrums correlate differently with the localization properties of the eigenstates depending on the strength of pairing and hopping range. 

\end{abstract}

\maketitle

\section{Introduction}
\label{sec1}

A theoretical model to explain the absence of diffusion in specific random lattice structures characterized by randomly distributed energy from one lattice site to another was formulated by P. W. Anderson in 1958 \cite{PhysRev.109.1492}. Such a system leads to the complete localization of all eigenfunctions. The Anderson localization has important implications for the transport of electrons in disordered materials. For 1D and 2D Anderson models, the scaling theory explains that even with a very weak disorder, all the states of the system transition from an ergodic or delocalized state to a fully localized state at the thermodynamic limit \cite{PhysRevLett.42.673, PhysRevLett.100.013906, PhysRevLett.105.163905}. However, a distinctive feature emerges in three-dimensional systems, where mobility edges or energy-dependent localization transitions appear \cite{ziman1969localization, PhysRevB.11.3697,bulka1985mobility}. These mobility edges serve as boundaries that separate localized and delocalized states, marking a fundamental departure from the behavior observed in lower dimensions.

In order to observe localization transition in the lower dimension, the 1D Aubry-Andr\'e-Harper (AAH) model was introduced. In this model, the onsite random disorder is replaced by a quasiperiodic cosine potential $V_n = V_0 \cos(2\pi\beta n + \theta)$, where $V_n$ is the potential at the $n$-th site of strength $V_0$, the parameter $\beta$ is irrational that decides the quasiperiodic nature of the potential, and $\theta$ is a phase factor. This model does not show a mobility edge if we restrict hopping in the system only to the nearest neighbor (NN). Here, all the single-particle energy eigenstates of the system made a sharp simultaneous localization transition \cite{AubryAndre,PhysRevA.80.021603, Wilkinson1984CriticalPO}. Specifically, they become exponentially localized once the disorder surpasses a critical threshold. Relaxing the restriction on the hopping only between the NN sites, we generalize the AAH model by allowing long-range hopping in the AAH model. This generalized AAH (GAAH) model exhibits mobility edge \cite{riklund1986extension, PhysRevA.75.063404, PhysRevLett.104.070601, PhysRevB.83.075105, PhysRevB.96.085119, PhysRevLett.123.025301, PhysRevB.103.075124, PhysRevLett.123.070405, 10.21468/SciPostPhys.12.1.027}. One dimensional GAAH model has been realized experimentally in ultracold atoms \cite{PhysRevLett.129.103401,PhysRevLett.126.040603}.

Recently, there has been a notable increase in exploring mobility edges in the GAAH model with power law hopping \cite{PhysRevLett.123.025301, PhysRevB.103.075124, PhysRevB.83.075105}. In this model, the hopping strength decays in power law fashion as $s^{-\xi}$, where $s$ is the separation between two sites and the index parameter $\xi$ can be tuned to set the hopping range. This system displays a transition from delocalized-to-multifractal states for long-range hopping, which is set by considering a small value of $\xi$. This transition is accompanied by the emergence of plateaus in the eigenstate spectrum, which depend on the quasiperiodic parameter $\beta$. A unique fraction of the delocalized states characterizes each plateau $P_l$, where the integer parameter $l$ denotes the plateau index. The value of this fraction is determined by the quasiperiodic parameter $\beta$ and the system size $N$ as $\beta^l N$. Additionally, the relationship between the fraction of delocalized eigenstates and a broader class of irrational Diophantine numbers, known as the metallic mean family, has been established within these intermediate regimes \cite{PhysRevB.103.075124}.

There has been a growing interest in non-Hermitian systems in recent years due to their distinctive characteristics such as complex energy spectra \cite{ashida2020non, bergholtz2021exceptional}, exceptional points \cite{wang2021topological, li2020symmetry, yuce2018pt, jin2017schrieffer, zhu2014pt, xu2020fate}, non-Hermitian skin effect \cite{PhysRevLett.124.056802, PhysRevLett.124.086801, PhysRevLett.125.126402} and the violation of conventional bulk-boundary correspondence \cite{yao2018edge, song2019non, lee2016anomalous}. All these characteristics have no Hermitian counterpart. Recently, investigation of mobility edges has been extended to various 1D tight-binding non-Hermitian models, including systems with power-law hopping \cite{PhysRevB.105.014207, PhysRevB.102.024205, PhysRevB.103.174205, Zhou2021TopologicalDT, PhysRevB.104.014202, PhysRevB.101.174205, PhysRevB.104.224204, PhysRevB.107.174205}. These studies have revealed interesting results, such as $\beta$-dependent robust $P_l$ regimes for weak non-Hermiticity. For stronger non-Hermiticity, the $P_l$ regimes disappear. Additionally, they found that localization properties, i.e., the long-range hopping-induced delocalized-to-multifractal edge and the short-range hopping-induced delocalized-to-localized edge, are robust against the non-Hermitian effect and are well-characterized by the fractal dimension $D_2$.

Recent research has revealed the profound impact of $p$-wave pairing on both Hermitian \cite{PhysRevLett.110.176403,PhysRevLett.110.146404,PhysRevB.106.024204,Zeng_2018,PhysRevB.100.064202,PhysRevB.100.064202,PhysRevB.103.104202, PhysRevA.105.013315,PhysRevB.93.104504} and non-Hermitian AAH model \cite{PhysRevA.95.062118,PhysRevB.103.214202,PhysRevB.108.014204}. The pairing term introduces significant changes in system dynamics, localization behaviors, and topological transition properties. In the study of the non-Hermitian AAH model, the incorporation of $p$-wave pairing introduces chiral (or particle-hole) symmetry to the model, which is a necessary condition for the presence of unpaired Majorana zero modes (MZM) \cite{PhysRevB.103.214202}.

Motivated by these ongoing investigations, our present study investigates the influence of pairing in the non-Hermitian GAAH (NH-GAAH) model with power-law hopping. Our primary interest is understanding the behavior of the $P_l$ regimes observed in the power-law hopping scenario within the NH-GAAH model with pairing. Furthermore, we present a comprehensive phase diagram illustrating the transitions associated with the localization and unconventional real-to-complex transitions within the system \cite{PhysRevB.103.104203}.

This paper is organized as follows. In the next section, Sec. \ref{sec2}, we introduce an NH-GAAH model that includes the power law hopping and NN superconducting pairing. Section \ref{sec3} discusses the characteristics of the central gap in the NH-GAAH chain with $p$-wave pairing for short- and long-range hopping. In section Sec. \ref{sec4}, we investigate the robustness of the $P_l$ regimes against pairing effects under weak non-Hermiticity conditions with long-range hopping in the system. The following section, Sec. \ref{sec5}, explores unconventional real-to-complex transition, considering both the hopping cases with pairing. Finally, we conclude our findings in Sec. \ref{sec6}.

\section{Model Hamiltonian} 
\label{sec2}

We investigate a 1D NH-GAAH system composed of long-range hopping with power-law decay, NN pairing, and complex onsite potentials. The Hamiltonian of the system can be expressed as follows: 
\begin{equation}
  \mathbf{
  \begin{split}
    H &= \sum_{j=0}^{N-1}\Bigl[ \sum_{s=1}^{N-1} \frac{-t}{s^\xi}( {c}_{j+s}^{\dagger}  {c}_j +  {c}_j^\dagger  {c}_{j+s}) \\
    &\quad - \mu f(j)(2  {c}_j^\dagger  {c}_j - 1) + \Delta ( {c}_{j+1}^\dagger  {c}_j^\dagger +  {c}_{j}  {c}_{j+1})\Bigr].
  \end{split}}
  \label{Hamiltonian}
\end{equation}
The non-Hermiticity in the Hamiltonian is introduced by setting the onsite potential $f(j)$ as a complex function 
\begin{equation}
f(j)= \cos(2\pi \beta j + \theta) + i h \sin(2\pi \beta j).
\end{equation}
Here, $h$ controls the strength of non-Hermiticity. This form of potential enables us to isolate and observe the sole effects of non-Hermiticity. In contrast, if we consider a widely used potential of the form $f(j) = \cos(2\pi \beta j + \theta + ih)$, then this potential introduces a Hermitian potential whose strength also varies with non-Hermitian parameter $h$ \cite{PhysRevLett.122.237601, PhysRevB.108.014204, PhysRevB.108.075121, PhysRevB.104.014202,PhysRevB.104.024201, PhysRevB.104.224204, PhysRevB.107.174205}. As mentioned earlier, the parameter $\beta$ decides the quasiperiodic nature of the onsite potential, and $\theta$ is the phase factor that can be used to shift the potential in space. Throughout this paper, we set $\theta = 0$. Other parameters are the hopping amplitude $t$, the NN superconducting pairing amplitude $\Delta \in \mathbbm{R}$, and the parameter $\xi$ is the power-law index. The operators $ {c}_j\, ( {c}_j^\dagger)$ are the fermionic annihilation (creation) operators at the $j$-th site of the chain.  

This model specifically upholds particle-hole symmetry, i.e., the Hamiltonian satisfies the condition $(\mathcal{PC})H(\mathcal{PC})^{-1} = -H$, where $\mathcal{P}$ and $\mathcal{C}$ are parity and charge-conjugation operators, respectively. This symmetry leads to the pair of energies ($E$, -$E^*$) in the spectrum. However, this system does not have ${\mathcal PT}$-symmetry \cite{bender1998real}. A more detailed discussion about the symmetry properties of the Hamiltonian is presented in Appendix \ref{appendixA}. Following a standard procedure, we express the Hamiltonian given in Eq. \eqref{Hamiltonian} in the Bogoliubov-de Gennes (BdG) basis: 
$\chi = (c_{0}, c_{0}^{\dagger}, \ldots, c_{N-1}, c_{N-1}^{\dagger})^{T}$ as:
\begin{equation}
H = \chi^{\dagger} H^{\rm BdG}\chi
\end{equation}  
where
\begin{equation}
H^{\rm BdG} = 
\begin{pmatrix}
A_0 & B & C_2 & \cdots & B^{\dagger} \\
B^{\dagger} & A_1 & B & \cdots & C_{N-2} \\
\vdots & \vdots & \vdots & \ddots & \vdots \\
C^{\dagger}_{N-2} & C^{\dagger}_{N-3} & C^{\dagger}_{N-4} & \cdots & B \\
B & C^{\dagger}_{N-2} & C^{\dagger}_{N-3} & \cdots & A_{N-1}. \\
\end{pmatrix}
\label{BdG_Ham}
\end{equation}
In the BdG Hamiltonian, at the extreme corners, $2 \times 2$ matrices $B$ and $B^\dagger$ appear for the PBC, while these matrices disappear for the open boundary conditions (OBC). The Hamiltonian $H^{\rm BdG}$ is a $2N \times 2N$ matrix with $A_j = -\mu f(j)\sigma_{z}$, $B = t\sigma_z - \Delta i \sigma_{y}$, and $C_s = \frac{t}{{d_{s}}^{\alpha}}\sigma_{z}$ are $2 \times 2$ dimensional matrices. Here, we replace $s$ in the Hamiltonian with $d_s = \min(s, N - s)$ while imposing PBC, whereas under OBC, we set $d_s = s$ and disregard the terms containing $c_{j}$ with $j>N-1$. It is important to note that we do not need to employ the BdG basis for the $\Delta = 0$ case.

\section{Central gap analysis: Majorana modes and Massive Dirac modes}
\label{sec3}

\begin{figure*}
    \centering
    \includegraphics[width=0.9\textwidth]{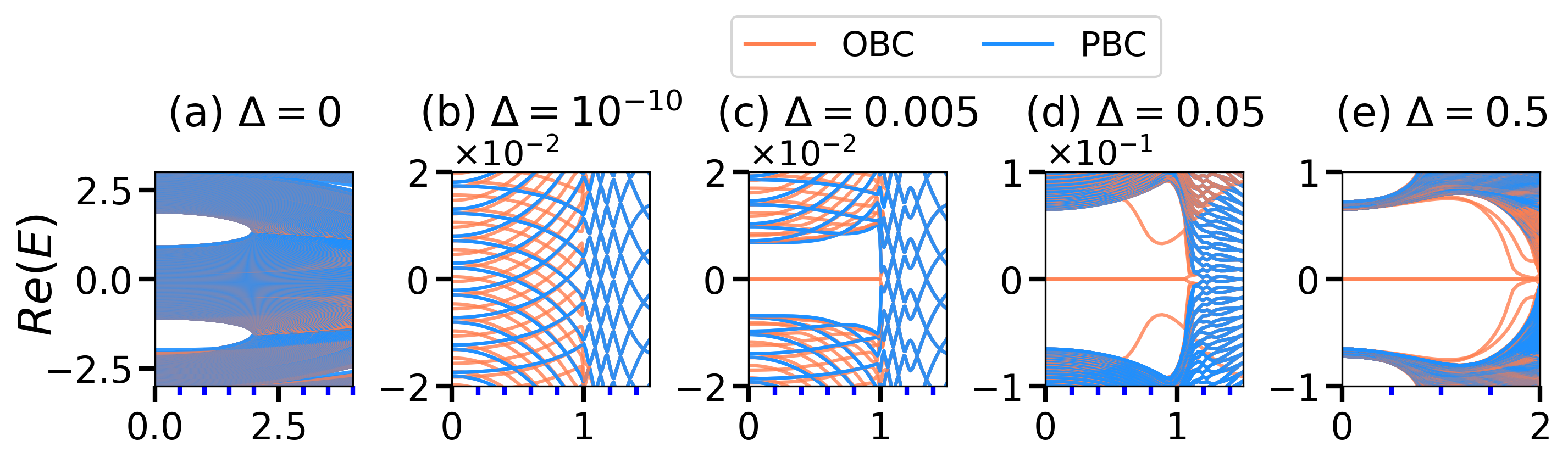}\\
    \vspace{-11pt} 
    \includegraphics[width=0.9\textwidth]{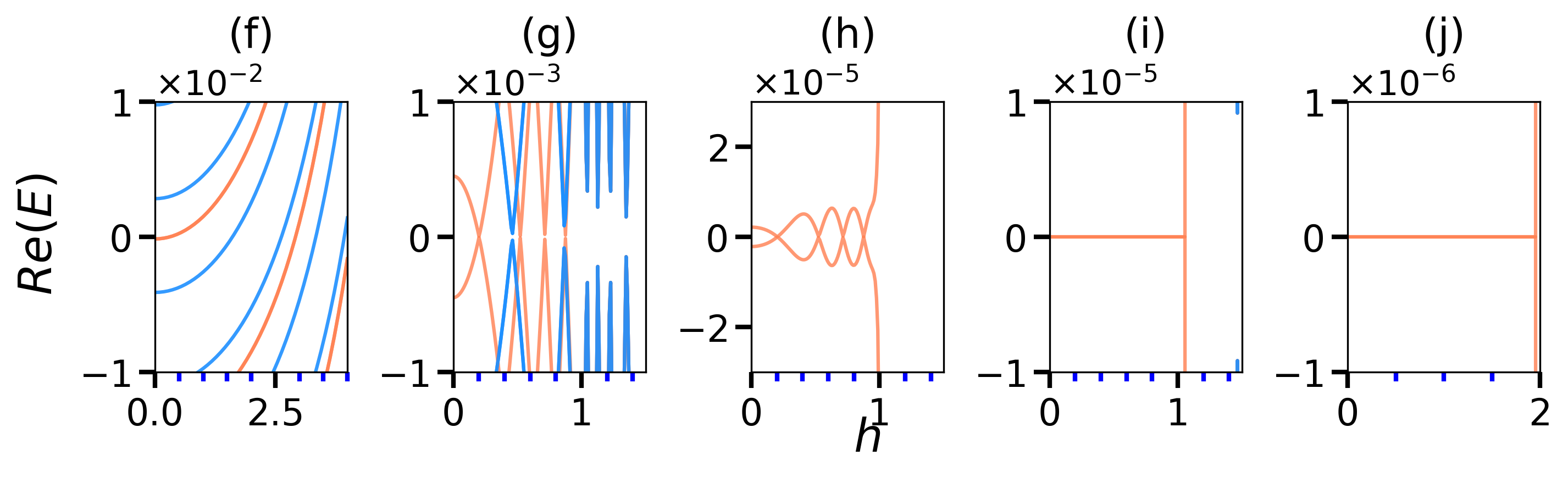}\\
    \includegraphics[width=0.9\textwidth]{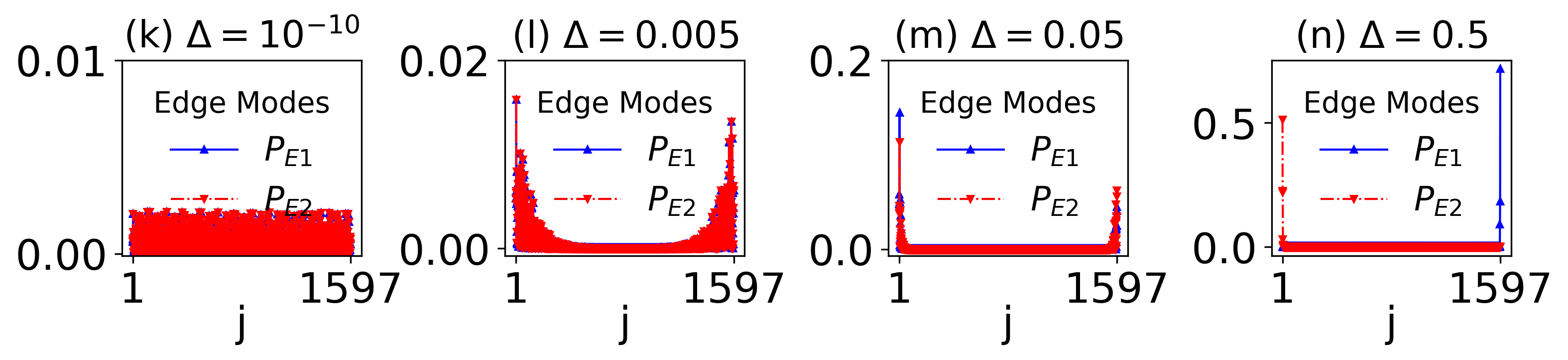}
    \caption{(a)-(e) Real parts of the eigenvalues are presented as a function of increasing pairing strength $\Delta$. (f)-(j) Zoomed-in views of the corresponding plots in (a)-(e), focusing the central gap region. (k)-(n) Two edge modes $P_E$  are highlighted in distinct colors: blue ($P_{E1}$) and red ($P_{E2}$). We observe that at $\Delta = 0$, there are no edge states, as shown in panels (a) and (f). For lower values of $\Delta$, such as $\Delta = 10^{-10}$ and $0.005$, the energy of the quasi-zero energy modes becomes oscillatory, resulting in oscillating hybridized states, as seen in panels (b) and (g) and in (c) and (h). However, panel (k) shows that for tiny pairing strength $\Delta = 10^{-10}$, the eigenstates corresponding to quasi-zero energy modes extend throughout the lattice, indicating that these are not edge-like states. In contrast, panel (l) demonstrates that the quasi-zero energy edge modes exhibit peaks at both edges. As the pairing strength $\Delta$ increases, the distance between these quasi-zero energy edge-like states grows, reducing their overlap, illustrated in panels (m) and (n). Consequently, the quasi-zero energy states begin to transition into zero-energy states, indicating localization at the edges.}
    \label{es_sr}
\end{figure*}

\begin{figure*}[htbp]
\begin{tabular}{cc}
    \centering
    \includegraphics[width=0.6\textwidth]{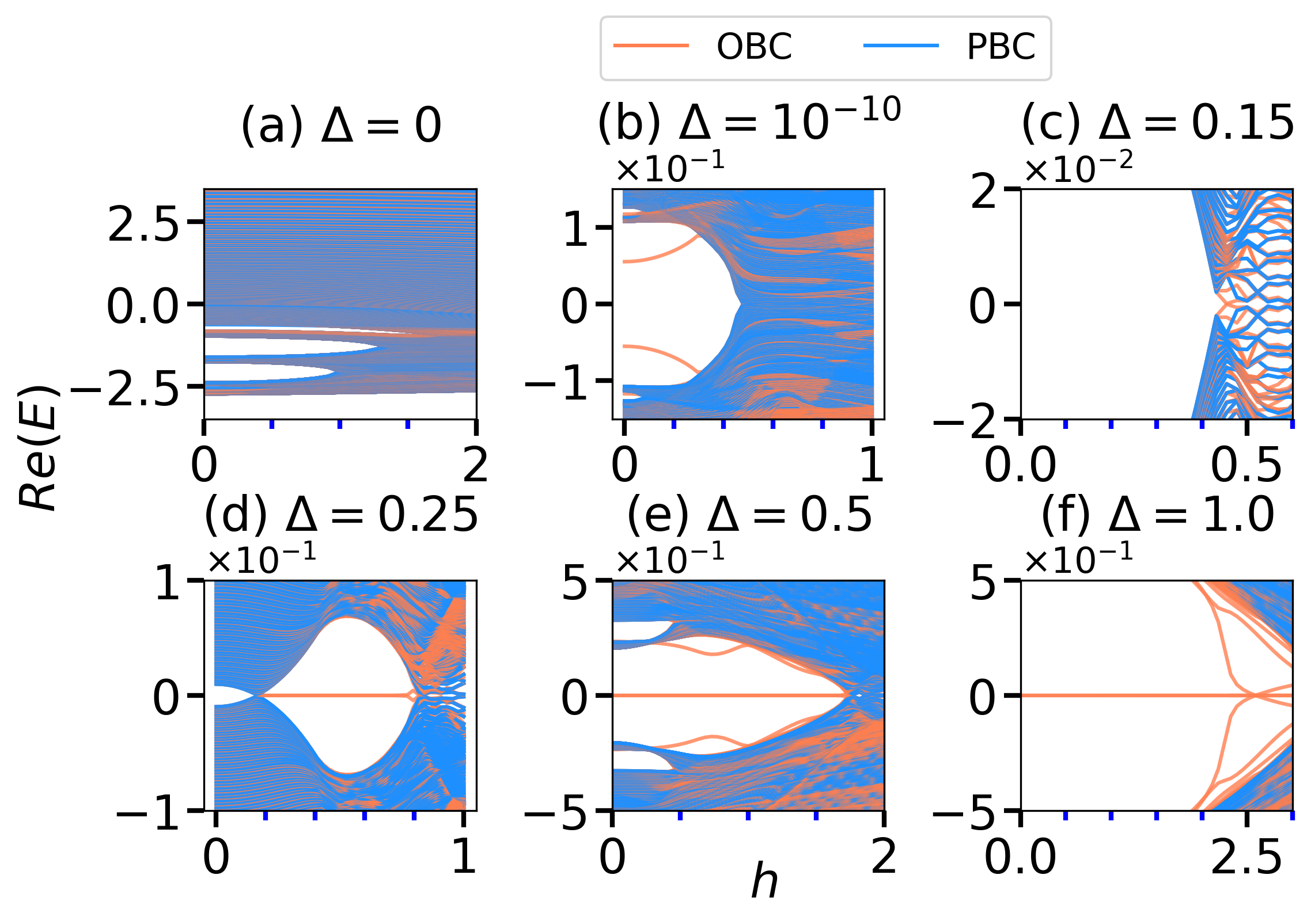}  &
    \vline{}
    \includegraphics[width=0.343\textwidth]{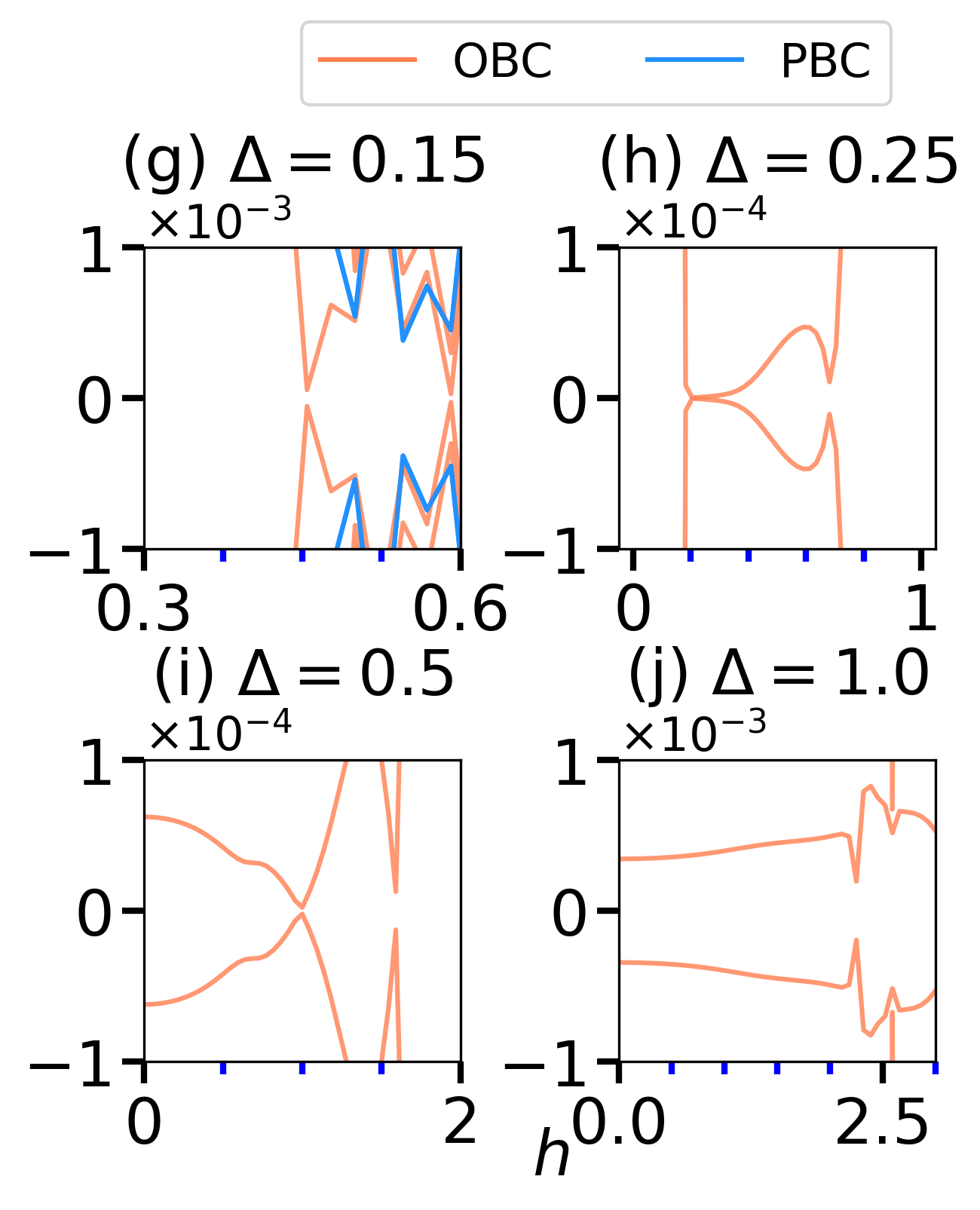} 
  \end{tabular}
   \includegraphics[width=0.9\textwidth]{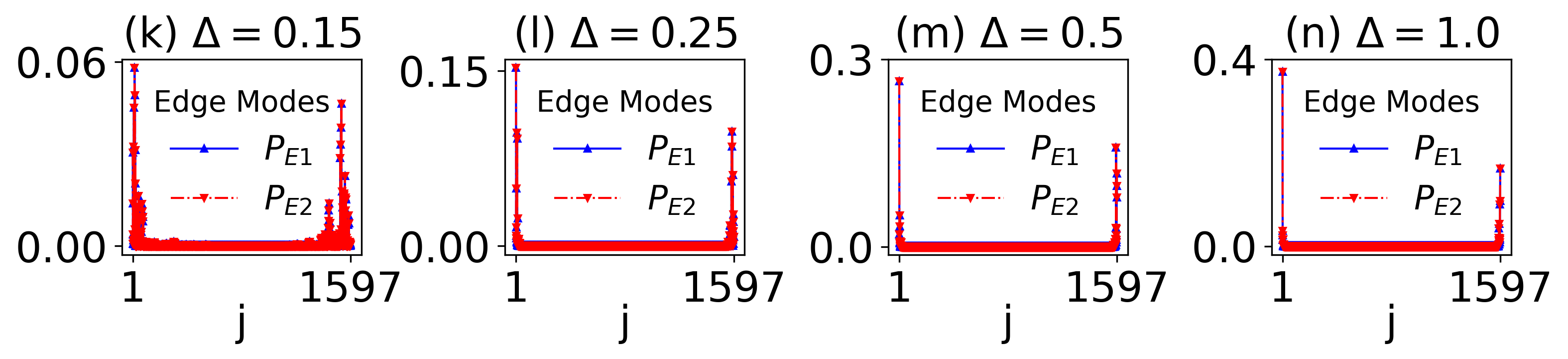} 
    \caption{(a)-(f) The real part of the energy spectrum is shown as a function of the non-Hermitian parameter $h$ for increasing values of the pairing strength $\Delta$ for the case of long-range hopping. We observe that for $\Delta = 0 \, , 10^{-10}$, there are no edge modes, as shown in panels (a) and (b). As the pairing strength $\Delta$ increases, the edge modes appear, depicted in panels (c)-(f). Zoomed-in views of the panels (c)-(f) around zero-energy are presented in panels (g)-(j). The edge modes for the same $\Delta$ values are shown in panels (k)-(n). The oscillatory edge modes appear for lower values of $\Delta$ as illustrated in panels (g)-(i) and in panels (k)-(m). The panels (j) and (n) show that the oscillatory nature decreases with increasing $\Delta$, i.e., the edge modes become more uniformly localized at the edges.}
    \label{es_lr}
\end{figure*}

This section concentrates on the system's characteristics near its central energy gap $E = 0$. More precisely, we study the effects of the short- and long-range hopping, considering the presence or absence of the pairing strength $\Delta$. Throughout this section, we exclusively concentrate on the real part of energy as a function of the non-Hermitian parameter $h$. Our system is comprised of $N$ sites. However, in the presence of a nonzero $\Delta$, we employ the BdG basis, which effectively doubles the basis size to $2N$. Here, we set the number of sites $N = 1597$ and keep it fixed throughout the paper. 

\subsection{Case I: Short-range hopping}

For the short-range hopping case, we set the power-law index $\xi = 5.0$. Figures \ref{es_sr} (a)-(e) illustrate the real part of eigenvalues plotted with respect to the non-Hermitian parameter $h$, corresponding to different values of $\Delta$. In addition, in Figs. \ref{es_sr} (f)-(j), we display an enlarged view of Figs. \ref{es_sr} (a)-(e) for the corresponding $\Delta$. In Figs. \ref{es_lr} (k)-(n), we present the edge modes with blue and red colors (denoted by $P_{E1}$ and $P_{E2}$ in the figures). In the absence of pairing, there are no zero-energy edge modes initially, as evident in Figs. \ref{es_sr}(a) and (f). 

However, with the introduction of even a tiny amount of pairing strength $\Delta = 10^{-10}$, we can see in Figs. \ref{es_sr}(b) and (g) that a symmetric energy spectrum appears about a central gap. This phenomenon is attributed to the inherent particle-hole symmetry in the system due to the superconducting pairing. For this tiny pairing strength and even for $\Delta = 0.005$, we observe an oscillating character of the quasi-zero energy modes [$\mathcal{O}(10^{-6})$] as shown in Figs. \ref{es_sr}(b) and (g), and in (c) and (h). Interestingly, this oscillating character of the eigenspectra is similar to that obtained for Majorana modes in the semiconductor-superconductor nanowire with finite length \cite{PhysRevB.97.155425, PhysRevB.87.094518}, tuned by external magnetic field and finite-size Su-Schrieffer-Heeger (SSH) chain with dissipative couplings \cite{PhysRevApplied.18.024038}. Here, the non-Hermitian parameter $h$ plays the role of the magnetic field in a nanowire for controlling Majorana oscillation. However, these oscillations are subtle and can only be observed in an enlarged view. As demonstrated in Ref. \cite{PhysRevApplied.18.024038}, the balanced gain and loss potential reduces the amplitude of the oscillations, making them difficult to detect. These oscillating edge modes accompany zero-energy crossings (refer Appendix \ref{appendixB}), which can have zero-energy conductance peak \cite{PhysRevB.97.155425}. Figure \ref{es_sr}(k) shows that the quasi-zero-energy modes corresponding to $\Delta = 10^{-10}$ are not edge states. For $\Delta = 0.005$, Fig. \ref{es_sr}(l) shows that both the quasi-zero-energy edge states have peaks at both edges. As the pairing strength is further increased to $\Delta = 0.05$ and $0.5$, Figs. \ref{es_sr}(i), (m), (j), and (n) show that the quasi-zero-energy edge modes smoothly transform into zero-energy edge modes. Particularly, for $\Delta = 0.5$, we observe that the zero energy edge modes are completely separated from each other (without any overlap) and localized at the two different edges of the lattice. This separation strongly suggests that these zero-energy edge states are MZM. Thus, for the short-range hopping case, on tuning the pairing strength, we observe a transition from the quasi-Majorana zero modes (qMZM), characterized by multiple crossings of the zero energy line, to MZM without any crossings.

\subsection{Case II: Long-range hopping}
 
We set the power law index $\xi = 0.2$. This smaller value relaxes the hopping range, and particles can directly hop from one site to a remote site. However, as we know, the hopping probability decreases in power-law fashion with distance between two sites. We observe the eigenspectra of the system as a function of the non-Hermitian parameter $h$ for increasing $\Delta$ values as shown in Figs. \ref{es_lr} (a)-(f). We do not observe central energy-gapped states for the pairing strengths $\Delta = 0$ and $10^{-10}$. Particularly for $\Delta = 0$, the energy spectrum exhibits a distinct lack of symmetry. This asymmetry in the spectrum is due to the presence of long-range hopping. However, as we introduce nonzero pairing with $\Delta = 10^{-10}$ in the system, we witness a noteworthy transformation in the spectrum. Even at this tiny value of $\Delta$, Fig. \ref{es_lr}(b) shows that a central gap appears in the spectrum about zero energy at $h = 0.45$. The system does not exhibit any distinct topological characteristics for these two values $\Delta = 0$ and $10^{-10}$. Therefore, we ignore these two cases for further studies. 

We now focus on the remaining four values of $\Delta = 0.15, \, 0.25, \, 0.5$ and $1.0$. For these values of $\Delta$, Figs. \ref{es_lr} (g)-(j) provide an enlarged view of the spectra around the central gap, which reveals the absence of any exact zero-energy modes in the system. In addition, we observe anti-crossing oscillation of these nonzero edge states for lower values of $\Delta$, i.e., for $\Delta < 1.0$ \cite{PhysRevB.97.155425}. For each of these values of $\Delta$, we select a couple of states from the central gap with the same value of $h$ such that their energy difference is minimum. We present those states in Figs. \ref{es_lr}(k)-(n). Interestingly, here we observe that, with the increment of $\Delta$, the magnitude of the energies of these selected states increase. However, they gradually localize with nonzero components at both ends of the system. The edge states' localization behavior is distinct from the MZM, but it is typical for the massive nonlocal edge states known as massive Dirac modes (MDM)  \cite{PhysRevLett.113.156402, Vodola_2016, PhysRevB.94.125121, PhysRevResearch.3.013148}. In these papers, MDM are observed in the topologically nontrivial system having long-range pairing with nearest-neighbor hopping. Here, we observe the same massive modes while studying the system, which has long-range hopping with nearest-neighbor pairing. Interestingly, the MDM also show oscillatory behavior for low values of $\Delta$.

\section{Plateau modulations by pairing interactions under weak non-Hermiticity}
\label{sec4}

The presence of plateaus in the quasiperiodic system with power-law hopping emphasizes a clear distinction between delocalized and multifractal states. These plateaus respond to changes in various system parameters, including non-Hermiticity \cite{PhysRevB.107.174205}. We quantify the effect of pairing on these plateaus by employing the conventional box-counting technique and investigate the fractal properties of the system. Here, we particularly study the (multi)fractal dimension of the energy eigenstates \cite{PhysRevLett.67.607}. Here, we set the number of sites in the system $N$. Consequently, the Hamiltonian of the system is a $N \times N$ matrix, and its eigenstates are $N$-dimensional complex vectors. The fractal dimension of the eigenstates is calculated in the following way: we divide the $N$ components of the eigenstates into $N_d$ boxes, and then each box will have $d=N/N_d$ number of components. If we expand the $i$-th eigenstate of the Hamiltonian $|\psi_i\rangle$ in the site basis as $\{|n\rangle,\, n = 1, \dots, N\}$, then $|\psi_i\rangle = \sum_n c_{in} |n\rangle$, where $c_{in}$ is the $n$-th component of the $i$-th eigenstate. The box probability for the $k$-th box of the $i$-th eigenstate 
\[p_k(d) = \sum\limits_{n = (k-1)d + 1}^{kd} |c_{in}|^2, ~{\rm where}~ k = 1, \dots, N_d\]
is defined as a suitable measure. Here, the occupation of the site $n$ for the $i$-th eigenstate is defined as $|c_{in}|^2 = |u_{i,n}|^2 + |v_{i,n}|^2$, where $\{u_{i,n},\, v_{i,n}\}$ are the coefficients of the eigenstates in the BdG basis. The generalized fractal dimension $D_q$ of any eigenstate can be calculated from the following relation:
\begin{equation}
D_q = \frac{1}{q-1}\lim_{d \rightarrow 0} \frac{\log\left(\sum\limits_{k=1}^{N_d} [p_k(d)]^q\right)}{\log d}.
\label{D_f_eq}
\end{equation}

\begin{figure*}
    \centering
    \includegraphics[width=0.92\textwidth]{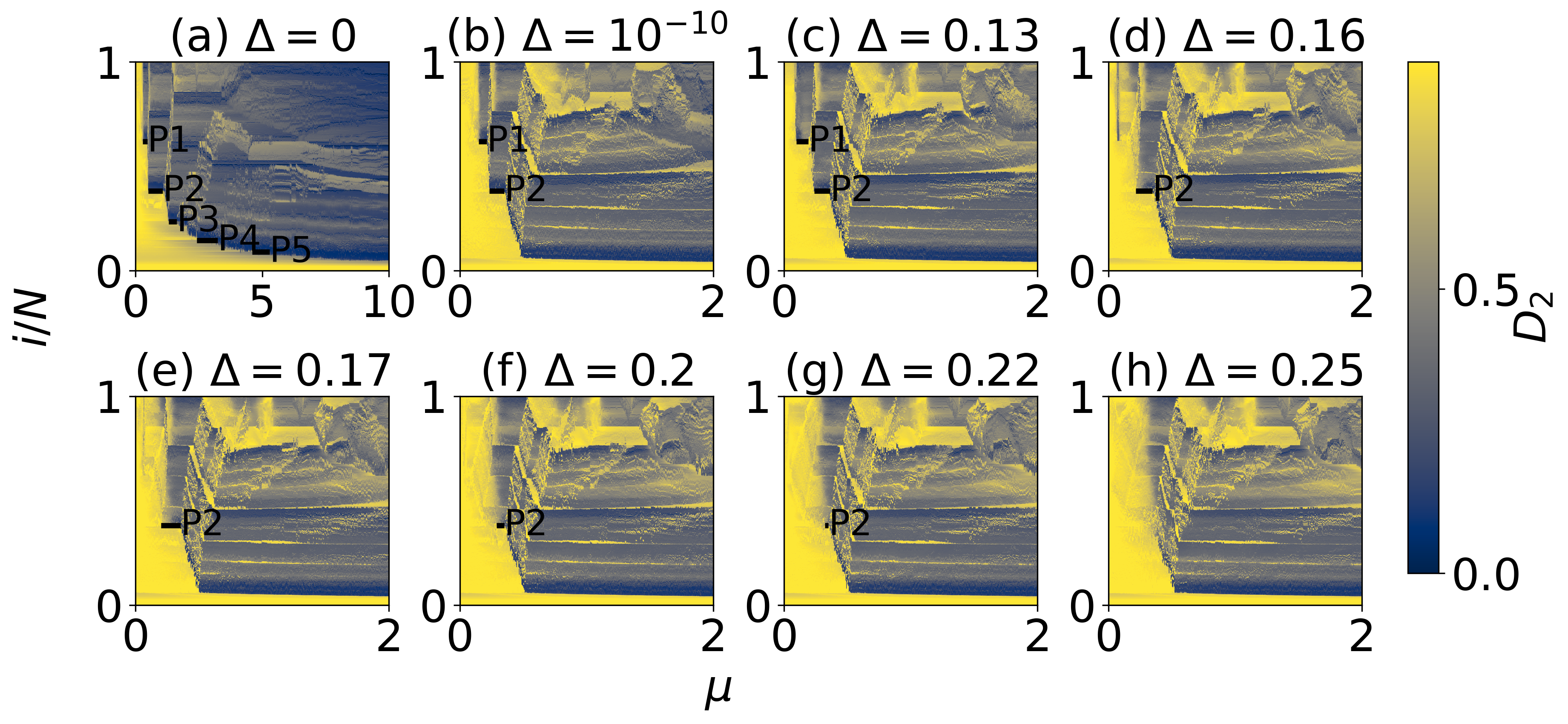}
    \caption{Fractal dimension $D_2$ (shown in color) of different eigenstates as a function of the chemical potential $\mu$ with a constant non-Hermitian parameter $h = 0.1$. (a) For $\Delta = 0$, five distinct plateaus are observed. Panels (b) and (c) show two distinct plateaus for $\Delta = 10^{-10}$ and also for $\Delta = 0.13$. As $\Delta$ reaches $0.16$, the first plateau diminishes, as shown in panel (d). Panel (e) exhibits that only a second plateau exists for $\Delta = 0.17$. Panels (f) and (g) display the narrowing down of the second plateau. Panel (h) depicts that no plateaus are discernible when $\Delta = 0.25$.}
    \label{Plateau_lr}
\end{figure*}

In this study, we observe that for the perfectly delocalized state, the generalized fractal dimension becomes almost independent of the scaling index $q$ and remains fixed around $D_q \simeq 1.0$. On the other hand, for the localized state, $D_q$ remains close to zero for all values of $q$. For the multifractal states, $D_q$ demonstrates a nonlinear dependence on the scaling index $q$ and remains within the range $0 < D_q < 1$. Like earlier, we set $N = 1597$ in this part of the study. We vary the box-counting index $d$ within $2$ to $20$. The fractal dimension $D_{q=2} = D_2$ is studied for all the energy eigenstates as a function of the potential strength $\mu$ for a fixed value of the non-Hermitian parameter $h=0.1$ and the power-law index $\xi = 0.2$. Note that, for this value of the power-law index, the system has long-range hopping. In Fig. \ref{Plateau_lr}, we explore the variation of $D_2$ of all the energy eigenstates as a function of the potential strength $\mu$ for different values of the pairing parameter $\Delta$. Here, we deliberately exclude the study of the short-range hopping cases because we do not observe any plateaus in these cases. 

In Fig. \ref{Plateau_lr}(a), we present the behavior of the fractal dimension $D_2$ for $\Delta = 0$. Here, we observe a noticeable decrease in the fraction of the number of delocalized eigenstates as $\mu$ increases. This observation suggests the presence of the well-known delocalized-to-multifractal (DM) edge in the eigenstate spectrum. The DM edge descends step-wise as the fraction of the delocalized eigenstates decreases with increasing $\mu$. In the decreasing DM edges region, we find several plateaus $P_l$ with $l = 1, \dots, 5$ appear. These plateaus are located at $\frac{i}{N} = \beta,\, \beta^2,\, \beta^3,\, \beta^4$, and $\beta^5$. Multifractality is demonstrated by plotting the average of $D_q$ of the eigenstates selected from the $P_2$ region with $\mu = 1.0$ as detailed in  Appendix \ref{appendixC}.

We then introduce very weak pairing $\Delta = 10^{-10}$ in the system. Figure \ref{Plateau_lr}(b) shows a sharp reduction in the number of distinct plateaus. We now observe that only two plateaus ($P_1$ and $P_2$) are survived. These plateaus correspond to the fractions $\frac{i}{N} = \beta$ and $\beta^2$, respectively. Figure \ref{Plateau_lr}(c) shows that these two plateaus survive till $\Delta = 0.13$. As we further increase to $\Delta = 0.16$, Fig. \ref{Plateau_lr}(d) shows that the first plateau $P_1$ begins to disappear, and only $P_2$ plateau remains prominent. With a tiny amount of increment in the pairing strength, we observe in Fig. \ref{Plateau_lr}(e) a complete disappearance of $P_1$ plateau at $\Delta = 0.17$, but $P_2$ plateau is still clearly visible. Further, on increasing pairing strength $\Delta$, we observe a narrowing down of the second plateau, as shown in Figs. \ref{Plateau_lr}(f) and (g). Figure \ref{Plateau_lr}(h) shows that, for $\Delta = 0.25$, all the plateaus disappear. This result implies that the characteristic plateaus, indicative of the long-range hopping, disappear as $\Delta$ increases.

\section{Unconventional real-to-complex and delocalized-to-multifractal transitions}
\label{sec5}

\begin{figure}
    \centering
    \includegraphics[width=0.5\textwidth]{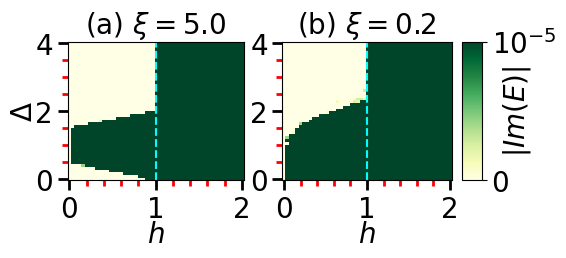} \\
     \includegraphics[width=0.48\textwidth]{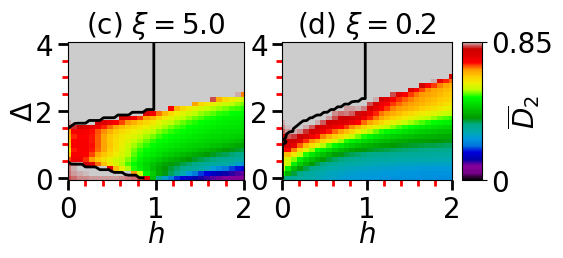}   
    \caption{The phase diagrams of the unconventional real-to-complex transition and delocalized-to-multifractal transition for different values of $\Delta$ and $h$. The largest value of the imaginary parts of the energy eigenvalues are presented to observe the real-to-complex transition for (a) $\xi = 5.0$ and (b) $\xi = 0.2$. Different color palettes distinguish the real and complex regions of the energies (see the color bar). In contrast, the mean fractal dimension $\overline{D}_2$ is presented for the same values of $\xi$ as in (c) and (d), respectively. The black solid line corresponds to the boundary of the real and the complex energy eigenvalues region.}
    \label{PT_DM}
\end{figure}

The NH-GAAH model with long-range hopping without superconducting $p$-wave pairing is a $\mathcal{PT}$ symmetric system \cite{PhysRevB.104.224204, PhysRevB.107.174205}. However, the version of the NH-GAAH model studied in this paper does not have $\mathcal{PT}$ symmetry due to the presence of the pairing term in the Hamiltonian. Interestingly, even without $\mathcal{PT}$ symmetry, superconducting models show an unconventional real-to-complex transition of the energy spectrum \cite{PhysRevB.103.104203, PhysRevB.110.094203}. Here, we delve into this unconventional transition in the spectrum and the transition from delocalized-to-multifractal states for short-range hopping with power-law index $\xi = 5.0$ and long-range hopping with $\xi = 0.2$. The phase diagrams presented in Figs. \ref{PT_DM}(a) and (b), demonstrate the variation of the maximum value of the imaginary part of the energy, denoted by $|{\rm Im}(E)|$, with the non-Hermitian parameter $h$ and the pairing strength $\Delta$. Figures \ref{PT_DM}(c) and (d) present another type of phase diagram where the mean fractal dimension $\overline{D}_2$ over all the eigenstates is also presented as a function of $h$ and $\Delta$.

Figure \ref{PT_DM}(a) presents the result for the short-range hopping case. For $h \lesssim 1.0$, we observe two windows of the region along $\Delta$-direction (represented by light-yellow color) where the energy spectrum is real-valued. The size of these windows decreases with increasing values of the non-Hermitian parameter $h$. This result implies that, for $h \lesssim 1.0$, the energy spectrum makes real-to-complex and then complex-to-real transitions as we increase the pairing strength from $\Delta = 0$ to larger values. Subsequently, for $h \geq 1.0$, we do not see any real-valued region of the energy spectrum for all values of the pairing strength $\Delta$. In Fig. \ref{PT_DM}(c), we present the behavior of the mean fractal dimension for the short-range hopping. Here, the black solid line represents the transition between real-to-complex energy eigenvalues. This figure shows that the delocalized region of the energy eigenstates overlaps with the real eigenvalue region. In contrast, the multifractal region of the eigenstates coincides with the complex region till $h \lesssim 1.0$. This outcome aligns with the findings presented in Ref. \cite{PhysRevB.103.214202}, where the NH-AAH model with $p$-wave pairing was examined, and the real-valued energy region coincided with the delocalized region. However, unlike the behavior of $|{\rm Im}(E)|$, we observe both multifractal and delocalized regions of the energy eigenstates along $\Delta$-direction when $h \geq 1.0$. 

In Figs. \ref{PT_DM}(b) and (d), we study the long-range hopping case with $\xi = 0.2$. Here, in $h \lesssim 1.0$ region, we observe only a single window along $\Delta$-direction [represented by light-yellow color in Fig. \ref{PT_DM}(b)] where the energy eigenvalues are real. Like the short-range case, here also, the window size decreases with the increase of the parameter $h$; and subsequently, when $h \geq 1.0$, the energy eigenvalues become complex for all values of the parameter $\Delta$. However, Fig. \ref{PT_DM}(d) shows that the multifractal region of the eigenstates is almost overlapping with the complex eigenvalues region for $h \lesssim 1.0$. For the long-range case, when $h \geq 1.0$, we again observe the presence of multifractal and delocalized eigenstates along $\Delta$-direction. This observation differs from the scenario where the long-range hopping was considered without pairing \cite{PhysRevB.107.174205}. In this work, the real-to-complex transition aligned with the delocalized-to-multifractal transition.

\section{Summary}
\label{sec6}

This study examines the influence of pairing on the NH-GAAH model, characterized by power-law hopping. Our investigation has revealed several key findings. For small pairing strength $\Delta$, we observe oscillating qMZM in the case of short-range hopping and oscillating MDM in the case of long-range hopping. The oscillatory nature of both decreases as the pairing strength increases, and the edge states become more and more localized. Additionally, we investigate the behavior of the plateaus in the fractal dimension of the eigenstates for the long-range hopping case with weak non-Hermiticity. As the pairing is introduced in the system, the number of plateaus decreases from $5$ to $2$. The plateaus gradually disappear as we increase the pairing strength $\Delta$. We also provide a phase diagram that outlines the transitions associated with localization and unconventional real-to-complex transitions within the system. Notably, in the case of short-range hopping,  the energy spectrum shows real-to-complex and complex-to-real transition as the pairing strength increases, particularly when $h \lesssim 1.0$. For this range of values of $h$, the real energies and the delocalized eigenstates on the $h-\Delta$ plane coincide. The spectrum is complex for $h \geq 1.0$, but the eigenstates can be delocalized or multifractal depending on the pairing strength $\Delta$. Unlike the short-range hopping case, for the long-range hopping, a single window along the pairing strength direction exists where the eigenvalues are real. Here, the delocalized states do not coincide with the real eigenvalues. Again, the eigenvalues are complex for $h \geq 1.0$, accompanied by delocalized or multifractal eigenstates depending again on the pairing strength $\Delta$. These findings collectively emphasize the interplay of the superconducting pairing and the power-law hopping in the NH-GAAH model.

\appendix

\section{The particle-hole $\mathcal{PC}$ and parity-time $\mathcal{PT}$ symmetry}
\label{appendixA}

The particle-hole and the parity-time symmetries are defined as: 
\[(\mathcal{PC})H(\mathcal{PC})^{-1} = -H,\]
\[(\mathcal{PT}) H (\mathcal{PT})^{-1} = H.\] 
Here, the parity (spatial reflection) operator is defined as \[\mathcal{P}^{-1} c_j \mathcal{P} = c_{N+1-j},\] and the charge conjugation operator is defined as follows 
\[\mathcal{C}c_j\mathcal{C}^{-1}= ic^\dagger_j ~{\rm and}~ \mathcal{C}i\mathcal{C}^{-1}= -i.\] 
Further, the time reversal operator $\mathcal{ {T}}$ is defined as \[\mathcal{T} i \mathcal{T}^{-1} = -i.\] 
On applying charge conjugation operation on the pairing term:
\begin{align}
    &\mathcal{C}\Delta ( {c}_{j+1}^\dagger  {c}_j^\dagger +  {c}_{j}  {c}_{j+1})\mathcal{( {C})}^{-1} \nonumber \\
    &= \Delta ((-i)  {c}_{j+1} (-i)  {c}_{j} + i {c}_{j}^\dagger   i  {c}_{j+1}^\dagger ) \nonumber \\
      &= \Delta (-  {c}_{j+1}  {c}_{j} - {c}_{j}^\dagger  {c}_{j+1}^\dagger ) \nonumber \\
            &= -\Delta (  {c}_{j+1}  {c}_{j}+ {c}_{j}^\dagger  {c}_{j+1}^\dagger ) 
\label{a1}
\end{align}
Further on applying parity operation to Eq. \eqref{a1}:
\begin{align}
    &\mathcal{( {P} {C})} \Delta ( {c}_{j+1}^\dagger  {c}_j^\dagger +  {c}_j  {c}_{j+1}) \mathcal{( {P} {C})}^{-1} \nonumber \\
    &= \mathcal{( {P})}-\Delta (  {c}_{j+1}  {c}_j +  {c}_j^\dagger  {c}_{j+1}^\dagger )\mathcal{( {P})}^{-1} \nonumber \\
        &= -\Delta ( {c}_{N+1-(j+1)}  {c}_{N+1-j}+  {c}_{N+1-j}^\dagger  {c}_{N+1-(j+1)}^\dagger) \nonumber \\
    &=-\Delta  ( {c}_{N-j}  {c}_{N+1-j}+  {c}_{N+1-j}^\dagger   {c}_{N-j}^\dagger ) \nonumber \\
    &= -\Delta ( {c}_{j}  {c}_{j+1} +  {c}_{j+1}^\dagger   {c}_{j}^\dagger ) ~[{\rm where},~ N-j = j]   \nonumber \\
    &= -\Delta ( {c}_{j+1}^\dagger  {c}_{j}^\dagger +  {c}_{j}  {c}_{j+1})
\label{a2}
\end{align}
Hence, the pairing term is symmetric under particle-hole operation. Similarly, the other terms of the Hamiltonian are also symmetric under this operation. However, when $\mathcal{PT}$ operator is applied to the pairing term of the Hamiltonian, the $\mathcal{PT}$ symmetry is violated, as shown below:
\begin{align}
    &\mathcal{( {P} {T})}\Delta ( {c}_{j+1}^\dagger  {c}_j^\dagger +  {c}_{j}  {c}_{j+1})\mathcal{( {P} {T})}^{-1} \nonumber \\
    &= \Delta ( {c}_{N+1-(j+1)}^\dagger  {c}_{N+1-j}^\dagger +  {c}_{N+1-j}  {c}_{N+1-(j+1)}) \nonumber \\
    &=\Delta  ( {c}_{N-j}^\dagger  {c}_{N+1-j}^\dagger +  {c}_{N+1-j}  {c}_{N-j}) \nonumber\\
    &= \Delta ( {c}_{j}^\dagger  {c}_{j+1}^\dagger +  {c}_{j+1}  {c}_{j}) ~[{\rm where},~ N-j = j] \nonumber \\
    &= -\Delta ( {c}_{j+1}^\dagger  {c}_{j}^\dagger +  {c}_{j}  {c}_{j+1}) 
\end{align}
For the other terms of the Hamiltonian, $\mathcal{PT}$ symmetry is not violated. It is violated due to the presence of a real-valued pairing term. 
Hence, it can be shown that the Hamiltonian has $\mathcal{PC}$ symmetry but lacks $\mathcal{PT}$ symmetry.

\section{Ground state fermion parity and Crossings}
\label{appendixB}

\begin{figure}[t]
\begin{tabular}{cc}
    \centering
    \includegraphics[width=0.24\textwidth]{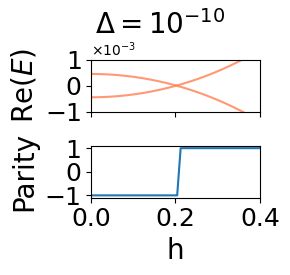} &
   \includegraphics[width=0.2\textwidth]{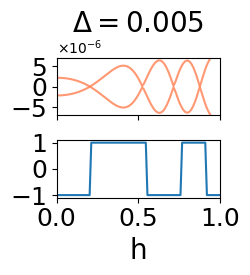}
\end{tabular}   
    \caption{The energy spectrum and ground state fermion parity for two values of the pairing strength: $\Delta = 10^{-10}$ and $\Delta = 0.005$. Notably, the fermion parity of the ground state undergoes a transition at $h=0.201$ for $\Delta = 10^{-10}$; and at $h=0.201$, $h=0.525$, $0.717$, and $0.878$ for $\Delta = 0.005$.}
  \label{parity}
\end{figure}

We observe the oscillations of the qMZM in the case of the short-range hopping with weak values of the pairing $\Delta$ (Figs. \ref{es_sr}(b), (g) and (c), (h). The genuine crossings and avoided ones are distinguished by investigating the behavior of the fermion parity of the ground state, as previously examined in Ref. \cite{PhysRevB.94.115166}. The ground state fermion parity is defined as:
\begin{equation}
P = {\rm sgn}[{\rm Pf}(H_M)]
\end{equation}
Here, Pf denotes the Pfaffian \cite{Wimmer}, while $H_M$ denotes the Hamiltonian in the Majorana basis with OBC. Employing the Majorana basis is pivotal due to its facilitation in expressing the Hamiltonian as a skew-symmetric matrix, a fundamental requirement for Pfaffian computation.

In Fig. \ref{parity}, we present the energy spectrum alongside the ground state fermion parity for $\Delta = 10^{-10}$ and $\Delta = 0.005$. At $h=0.201$, a single crossing is observed within the energy gap for $\Delta = 10^{-10}$. However, for $\Delta = 0.005$, four crossings manifest at $h=0.201$, $0.525$, $0.717$, and $0.878$. Remarkably, the fermion parity undergoes a sign change at these distinct crossing points. This behavior confirms that, at the crossing points, the exchange of the two orthogonal modes occurs, affirming the genuine nature of the crossings rather than the avoided one. Thus, these oscillating qMZM are accompanied by discrete zero-energy crossings.

\section{Multifractal properties}
\label{appendixC}

\begin{figure}[ht]
    \centering
    \includegraphics[width=0.45\textwidth]{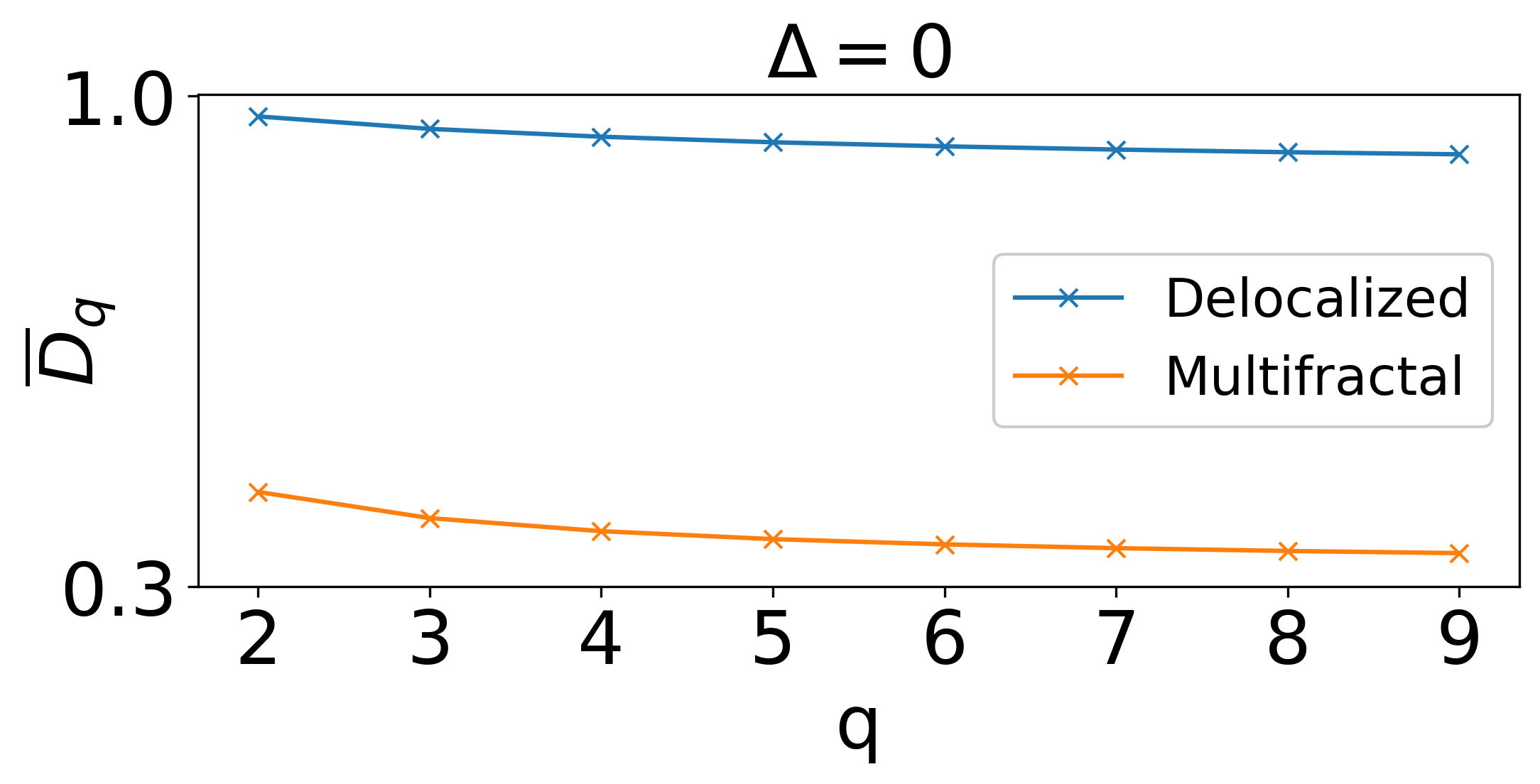}
    \caption{The average fractal dimension $\overline{D}_q$ as a function of $q$ is presented for $\Delta = 0$, with $\mu = 1.0$, selected from the $P_2$ regime. Here, we set the system size $N = 1597$. Notably, $\overline{D}_q$ tends toward unity for the lowest $\beta^2 N$ eigenstates, signifying their delocalized nature. Conversely, $\overline{D}_q$ exhibits variation with $q$ for the highest $(1-\beta^2)N$ eigenstates, illustrating their multifractal characteristics.}
    \label{Plateausfractal}
\end{figure}

In Fig. \ref{Plateausfractal}, the presence of the multifractality in the eigenstates of the system is presented by calculating the average of $D_q$ of the eigenstates selected from the $P_2$ region with $\mu = 1.0$. Here, we have calculated two averages of the multifractal dimension, denoted by $\overline{D}_q$. (I) average over the eigenstates below the $P_2$ plateau (marked by yellow region), where the states are highly delocalized; and (II) above the plateau $P_2$ (marked by darker region), where the states are primarily multifractal. The results corresponding to (I) are shown in the figure by solid blue lines with system size $N = 1597$. The solid orange line in the figure shows the result for (II) for the same system size. This result remains invariant even for larger system sizes $N = 2584,\, 4181$, and $6765$ (following three Fibonacci numbers greater than $1597$). In the case of (I), we find that $\overline{D}_q$ is almost insensitive to the scaling index $q$ and remains close to {\it unity}. This result suggests that the eigenstates are delocalized in this region. On the other hand, for case (II), we see that $\overline{D}_q$ is sensitive to $q$ and varies nonlinearly as a function of $q$. This result indicates the multifractal nature of the eigenstates in this region.  


\begin{thebibliography}{70}%
\makeatletter
\providecommand \@ifxundefined [1]{%
 \@ifx{#1\undefined}
}%
\providecommand \@ifnum [1]{%
 \ifnum #1\expandafter \@firstoftwo
 \else \expandafter \@secondoftwo
 \fi
}%
\providecommand \@ifx [1]{%
 \ifx #1\expandafter \@firstoftwo
 \else \expandafter \@secondoftwo
 \fi
}%
\providecommand \natexlab [1]{#1}%
\providecommand \enquote  [1]{``#1''}%
\providecommand \bibnamefont  [1]{#1}%
\providecommand \bibfnamefont [1]{#1}%
\providecommand \citenamefont [1]{#1}%
\providecommand \href@noop [0]{\@secondoftwo}%
\providecommand \href [0]{\begingroup \@sanitize@url \@href}%
\providecommand \@href[1]{\@@startlink{#1}\@@href}%
\providecommand \@@href[1]{\endgroup#1\@@endlink}%
\providecommand \@sanitize@url [0]{\catcode `\\12\catcode `\$12\catcode
  `\&12\catcode `\#12\catcode `\^12\catcode `\_12\catcode `\%12\relax}%
\providecommand \@@startlink[1]{}%
\providecommand \@@endlink[0]{}%
\providecommand \url  [0]{\begingroup\@sanitize@url \@url }%
\providecommand \@url [1]{\endgroup\@href {#1}{\urlprefix }}%
\providecommand \urlprefix  [0]{URL }%
\providecommand \Eprint [0]{\href }%
\providecommand \doibase [0]{https://doi.org/}%
\providecommand \selectlanguage [0]{\@gobble}%
\providecommand \bibinfo  [0]{\@secondoftwo}%
\providecommand \bibfield  [0]{\@secondoftwo}%
\providecommand \translation [1]{[#1]}%
\providecommand \BibitemOpen [0]{}%
\providecommand \bibitemStop [0]{}%
\providecommand \bibitemNoStop [0]{.\EOS\space}%
\providecommand \EOS [0]{\spacefactor3000\relax}%
\providecommand \BibitemShut  [1]{\csname bibitem#1\endcsname}%
\let\auto@bib@innerbib\@empty
\bibitem [{\citenamefont {Anderson}(1958)}]{PhysRev.109.1492}%
  \BibitemOpen
  \bibfield  {author} {\bibinfo {author} {\bibfnamefont {P.~W.}\ \bibnamefont
  {Anderson}},\ }\bibfield  {title} {\bibinfo {title} {Absence of diffusion in
  certain random lattices},\ }\href {https://doi.org/10.1103/PhysRev.109.1492}
  {\bibfield  {journal} {\bibinfo  {journal} {Phys. Rev.}\ }\textbf {\bibinfo
  {volume} {109}},\ \bibinfo {pages} {1492} (\bibinfo {year}
  {1958})}\BibitemShut {NoStop}%
\bibitem [{\citenamefont {Abrahams}\ \emph {et~al.}(1979)\citenamefont
  {Abrahams}, \citenamefont {Anderson}, \citenamefont {Licciardello},\ and\
  \citenamefont {Ramakrishnan}}]{PhysRevLett.42.673}%
  \BibitemOpen
  \bibfield  {author} {\bibinfo {author} {\bibfnamefont {E.}~\bibnamefont
  {Abrahams}}, \bibinfo {author} {\bibfnamefont {P.~W.}\ \bibnamefont
  {Anderson}}, \bibinfo {author} {\bibfnamefont {D.~C.}\ \bibnamefont
  {Licciardello}},\ and\ \bibinfo {author} {\bibfnamefont {T.~V.}\ \bibnamefont
  {Ramakrishnan}},\ }\bibfield  {title} {\bibinfo {title} {Scaling theory of
  localization: Absence of quantum diffusion in two dimensions},\ }\href
  {https://doi.org/10.1103/PhysRevLett.42.673} {\bibfield  {journal} {\bibinfo
  {journal} {Phys. Rev. Lett.}\ }\textbf {\bibinfo {volume} {42}},\ \bibinfo
  {pages} {673} (\bibinfo {year} {1979})}\BibitemShut {NoStop}%
\bibitem [{\citenamefont {Lahini}\ \emph {et~al.}(2008)\citenamefont {Lahini},
  \citenamefont {Avidan}, \citenamefont {Pozzi}, \citenamefont {Sorel},
  \citenamefont {Morandotti}, \citenamefont {Christodoulides},\ and\
  \citenamefont {Silberberg}}]{PhysRevLett.100.013906}%
  \BibitemOpen
  \bibfield  {author} {\bibinfo {author} {\bibfnamefont {Y.}~\bibnamefont
  {Lahini}}, \bibinfo {author} {\bibfnamefont {A.}~\bibnamefont {Avidan}},
  \bibinfo {author} {\bibfnamefont {F.}~\bibnamefont {Pozzi}}, \bibinfo
  {author} {\bibfnamefont {M.}~\bibnamefont {Sorel}}, \bibinfo {author}
  {\bibfnamefont {R.}~\bibnamefont {Morandotti}}, \bibinfo {author}
  {\bibfnamefont {D.~N.}\ \bibnamefont {Christodoulides}},\ and\ \bibinfo
  {author} {\bibfnamefont {Y.}~\bibnamefont {Silberberg}},\ }\bibfield  {title}
  {\bibinfo {title} {Anderson localization and nonlinearity in one-dimensional
  disordered photonic lattices},\ }\href
  {https://doi.org/10.1103/PhysRevLett.100.013906} {\bibfield  {journal}
  {\bibinfo  {journal} {Phys. Rev. Lett.}\ }\textbf {\bibinfo {volume} {100}},\
  \bibinfo {pages} {013906} (\bibinfo {year} {2008})}\BibitemShut {NoStop}%
\bibitem [{\citenamefont {Lahini}\ \emph {et~al.}(2010)\citenamefont {Lahini},
  \citenamefont {Bromberg}, \citenamefont {Christodoulides},\ and\
  \citenamefont {Silberberg}}]{PhysRevLett.105.163905}%
  \BibitemOpen
  \bibfield  {author} {\bibinfo {author} {\bibfnamefont {Y.}~\bibnamefont
  {Lahini}}, \bibinfo {author} {\bibfnamefont {Y.}~\bibnamefont {Bromberg}},
  \bibinfo {author} {\bibfnamefont {D.~N.}\ \bibnamefont {Christodoulides}},\
  and\ \bibinfo {author} {\bibfnamefont {Y.}~\bibnamefont {Silberberg}},\
  }\bibfield  {title} {\bibinfo {title} {Quantum correlations in two-particle
  anderson localization},\ }\href
  {https://doi.org/10.1103/PhysRevLett.105.163905} {\bibfield  {journal}
  {\bibinfo  {journal} {Phys. Rev. Lett.}\ }\textbf {\bibinfo {volume} {105}},\
  \bibinfo {pages} {163905} (\bibinfo {year} {2010})}\BibitemShut {NoStop}%
\bibitem [{\citenamefont {Ziman}(1969)}]{ziman1969localization}%
  \BibitemOpen
  \bibfield  {author} {\bibinfo {author} {\bibfnamefont {J.~M.}\ \bibnamefont
  {Ziman}},\ }\bibfield  {title} {\bibinfo {title} {Localization of electrons
  in ordered and disordered systems ii. bound bands},\ }\href@noop {}
  {\bibfield  {journal} {\bibinfo  {journal} {Journal of Physics C: Solid State
  Physics}\ }\textbf {\bibinfo {volume} {2}},\ \bibinfo {pages} {1230}
  (\bibinfo {year} {1969})}\BibitemShut {NoStop}%
\bibitem [{\citenamefont {Licciardello}\ and\ \citenamefont
  {Economou}(1975)}]{PhysRevB.11.3697}%
  \BibitemOpen
  \bibfield  {author} {\bibinfo {author} {\bibfnamefont {D.~C.}\ \bibnamefont
  {Licciardello}}\ and\ \bibinfo {author} {\bibfnamefont {E.~N.}\ \bibnamefont
  {Economou}},\ }\bibfield  {title} {\bibinfo {title} {Study of localization in
  anderson's model for random lattices},\ }\href
  {https://doi.org/10.1103/PhysRevB.11.3697} {\bibfield  {journal} {\bibinfo
  {journal} {Phys. Rev. B}\ }\textbf {\bibinfo {volume} {11}},\ \bibinfo
  {pages} {3697} (\bibinfo {year} {1975})}\BibitemShut {NoStop}%
\bibitem [{\citenamefont {Bulka}\ \emph {et~al.}(1985)\citenamefont {Bulka},
  \citenamefont {Kramer},\ and\ \citenamefont {MacKinnon}}]{bulka1985mobility}%
  \BibitemOpen
  \bibfield  {author} {\bibinfo {author} {\bibfnamefont {B.}~\bibnamefont
  {Bulka}}, \bibinfo {author} {\bibfnamefont {B.}~\bibnamefont {Kramer}},\ and\
  \bibinfo {author} {\bibfnamefont {A.}~\bibnamefont {MacKinnon}},\ }\bibfield
  {title} {\bibinfo {title} {Mobility edge in the three dimensional anderson
  model},\ }\href {https://doi.org/10.1007/BF01312638} {\bibfield  {journal}
  {\bibinfo  {journal} {Zeitschrift fur Physik B Condensed Matter}\ }\textbf
  {\bibinfo {volume} {60}},\ \bibinfo {pages} {13} (\bibinfo {year}
  {1985})}\BibitemShut {NoStop}%
\bibitem [{\citenamefont {Serge~Aubry}(1980)}]{AubryAndre}%
  \BibitemOpen
  \bibfield  {author} {\bibinfo {author} {\bibfnamefont {G.~A.}\ \bibnamefont
  {Serge~Aubry}},\ }\bibfield  {title} {\bibinfo {title} {Analyticity breaking
  and anderson localization in incommensurate lattices},\ }\bibfield  {journal}
  {\bibinfo  {journal} {Annals of the Israel Physical society}\ }\href
  {https://doi.org///chaos.if.uj.edu.pl/~delande/Lectures/files/An.Is.Phys.Soc.pdf}
  {//chaos.if.uj.edu.pl/~delande/Lectures/files/An.Is.Phys.Soc.pdf} (\bibinfo
  {year} {1980})\BibitemShut {NoStop}%
\bibitem [{\citenamefont {Biddle}\ \emph {et~al.}(2009)\citenamefont {Biddle},
  \citenamefont {Wang}, \citenamefont {Priour},\ and\ \citenamefont
  {Das~Sarma}}]{PhysRevA.80.021603}%
  \BibitemOpen
  \bibfield  {author} {\bibinfo {author} {\bibfnamefont {J.}~\bibnamefont
  {Biddle}}, \bibinfo {author} {\bibfnamefont {B.}~\bibnamefont {Wang}},
  \bibinfo {author} {\bibfnamefont {D.~J.}\ \bibnamefont {Priour}},\ and\
  \bibinfo {author} {\bibfnamefont {S.}~\bibnamefont {Das~Sarma}},\ }\bibfield
  {title} {\bibinfo {title} {Localization in one-dimensional incommensurate
  lattices beyond the aubry-andr\'e model},\ }\href
  {https://doi.org/10.1103/PhysRevA.80.021603} {\bibfield  {journal} {\bibinfo
  {journal} {Phys. Rev. A}\ }\textbf {\bibinfo {volume} {80}},\ \bibinfo
  {pages} {021603} (\bibinfo {year} {2009})}\BibitemShut {NoStop}%
\bibitem [{\citenamefont {Wilkinson}(1984)}]{Wilkinson1984CriticalPO}%
  \BibitemOpen
  \bibfield  {author} {\bibinfo {author} {\bibfnamefont {M.}~\bibnamefont
  {Wilkinson}},\ }\bibfield  {title} {\bibinfo {title} {Critical properties of
  electron eigenstates in incommensurate systems},\ }\href
  {https://api.semanticscholar.org/CorpusID:122524806} {\bibfield  {journal}
  {\bibinfo  {journal} {Proceedings of the Royal Society of London. A.
  Mathematical and Physical Sciences}\ }\textbf {\bibinfo {volume} {391}},\
  \bibinfo {pages} {305} (\bibinfo {year} {1984})}\BibitemShut {NoStop}%
\bibitem [{\citenamefont {Riklund}\ \emph {et~al.}(1986)\citenamefont
  {Riklund}, \citenamefont {Liu}, \citenamefont {Wahlstrom},\ and\
  \citenamefont {Zhao-bo}}]{riklund1986extension}%
  \BibitemOpen
  \bibfield  {author} {\bibinfo {author} {\bibfnamefont {R.}~\bibnamefont
  {Riklund}}, \bibinfo {author} {\bibfnamefont {Y.}~\bibnamefont {Liu}},
  \bibinfo {author} {\bibfnamefont {G.}~\bibnamefont {Wahlstrom}},\ and\
  \bibinfo {author} {\bibfnamefont {Z.}~\bibnamefont {Zhao-bo}},\ }\bibfield
  {title} {\bibinfo {title} {Extension of the hopping range in incommensurate
  systems can generate localised states},\ }\href
  {https://doi.org/10.1088/0022-3719/19/30/004} {\bibfield  {journal} {\bibinfo
   {journal} {Journal of Physics C: Solid State Physics}\ }\textbf {\bibinfo
  {volume} {19}},\ \bibinfo {pages} {L705} (\bibinfo {year}
  {1986})}\BibitemShut {NoStop}%
\bibitem [{\citenamefont {Boers}\ \emph {et~al.}(2007)\citenamefont {Boers},
  \citenamefont {Goedeke}, \citenamefont {Hinrichs},\ and\ \citenamefont
  {Holthaus}}]{PhysRevA.75.063404}%
  \BibitemOpen
  \bibfield  {author} {\bibinfo {author} {\bibfnamefont {D.~J.}\ \bibnamefont
  {Boers}}, \bibinfo {author} {\bibfnamefont {B.}~\bibnamefont {Goedeke}},
  \bibinfo {author} {\bibfnamefont {D.}~\bibnamefont {Hinrichs}},\ and\
  \bibinfo {author} {\bibfnamefont {M.}~\bibnamefont {Holthaus}},\ }\bibfield
  {title} {\bibinfo {title} {Mobility edges in bichromatic optical lattices},\
  }\href {https://doi.org/10.1103/PhysRevA.75.063404} {\bibfield  {journal}
  {\bibinfo  {journal} {Phys. Rev. A}\ }\textbf {\bibinfo {volume} {75}},\
  \bibinfo {pages} {063404} (\bibinfo {year} {2007})}\BibitemShut {NoStop}%
\bibitem [{\citenamefont {Biddle}\ and\ \citenamefont
  {Das~Sarma}(2010)}]{PhysRevLett.104.070601}%
  \BibitemOpen
  \bibfield  {author} {\bibinfo {author} {\bibfnamefont {J.}~\bibnamefont
  {Biddle}}\ and\ \bibinfo {author} {\bibfnamefont {S.}~\bibnamefont
  {Das~Sarma}},\ }\bibfield  {title} {\bibinfo {title} {Predicted mobility
  edges in one-dimensional incommensurate optical lattices: An exactly solvable
  model of anderson localization},\ }\href
  {https://doi.org/10.1103/PhysRevLett.104.070601} {\bibfield  {journal}
  {\bibinfo  {journal} {Phys. Rev. Lett.}\ }\textbf {\bibinfo {volume} {104}},\
  \bibinfo {pages} {070601} (\bibinfo {year} {2010})}\BibitemShut {NoStop}%
\bibitem [{\citenamefont {Biddle}\ \emph {et~al.}(2011)\citenamefont {Biddle},
  \citenamefont {Priour}, \citenamefont {Wang},\ and\ \citenamefont
  {Das~Sarma}}]{PhysRevB.83.075105}%
  \BibitemOpen
  \bibfield  {author} {\bibinfo {author} {\bibfnamefont {J.}~\bibnamefont
  {Biddle}}, \bibinfo {author} {\bibfnamefont {D.~J.}\ \bibnamefont {Priour}},
  \bibinfo {author} {\bibfnamefont {B.}~\bibnamefont {Wang}},\ and\ \bibinfo
  {author} {\bibfnamefont {S.}~\bibnamefont {Das~Sarma}},\ }\bibfield  {title}
  {\bibinfo {title} {Localization in one-dimensional lattices with
  non-nearest-neighbor hopping: Generalized anderson and aubry-andr\'e
  models},\ }\href {https://doi.org/10.1103/PhysRevB.83.075105} {\bibfield
  {journal} {\bibinfo  {journal} {Phys. Rev. B}\ }\textbf {\bibinfo {volume}
  {83}},\ \bibinfo {pages} {075105} (\bibinfo {year} {2011})}\BibitemShut
  {NoStop}%
\bibitem [{\citenamefont {Li}\ \emph {et~al.}(2017)\citenamefont {Li},
  \citenamefont {Li},\ and\ \citenamefont {Das~Sarma}}]{PhysRevB.96.085119}%
  \BibitemOpen
  \bibfield  {author} {\bibinfo {author} {\bibfnamefont {X.}~\bibnamefont
  {Li}}, \bibinfo {author} {\bibfnamefont {X.}~\bibnamefont {Li}},\ and\
  \bibinfo {author} {\bibfnamefont {S.}~\bibnamefont {Das~Sarma}},\ }\bibfield
  {title} {\bibinfo {title} {Mobility edges in one-dimensional bichromatic
  incommensurate potentials},\ }\href
  {https://doi.org/10.1103/PhysRevB.96.085119} {\bibfield  {journal} {\bibinfo
  {journal} {Phys. Rev. B}\ }\textbf {\bibinfo {volume} {96}},\ \bibinfo
  {pages} {085119} (\bibinfo {year} {2017})}\BibitemShut {NoStop}%
\bibitem [{\citenamefont {Deng}\ \emph {et~al.}(2019)\citenamefont {Deng},
  \citenamefont {Ray}, \citenamefont {Sinha}, \citenamefont {Shlyapnikov},\
  and\ \citenamefont {Santos}}]{PhysRevLett.123.025301}%
  \BibitemOpen
  \bibfield  {author} {\bibinfo {author} {\bibfnamefont {X.}~\bibnamefont
  {Deng}}, \bibinfo {author} {\bibfnamefont {S.}~\bibnamefont {Ray}}, \bibinfo
  {author} {\bibfnamefont {S.}~\bibnamefont {Sinha}}, \bibinfo {author}
  {\bibfnamefont {G.~V.}\ \bibnamefont {Shlyapnikov}},\ and\ \bibinfo {author}
  {\bibfnamefont {L.}~\bibnamefont {Santos}},\ }\bibfield  {title} {\bibinfo
  {title} {One-dimensional quasicrystals with power-law hopping},\ }\href
  {https://doi.org/10.1103/PhysRevLett.123.025301} {\bibfield  {journal}
  {\bibinfo  {journal} {Phys. Rev. Lett.}\ }\textbf {\bibinfo {volume} {123}},\
  \bibinfo {pages} {025301} (\bibinfo {year} {2019})}\BibitemShut {NoStop}%
\bibitem [{\citenamefont {Roy}\ and\ \citenamefont
  {Sharma}(2021)}]{PhysRevB.103.075124}%
  \BibitemOpen
  \bibfield  {author} {\bibinfo {author} {\bibfnamefont {N.}~\bibnamefont
  {Roy}}\ and\ \bibinfo {author} {\bibfnamefont {A.}~\bibnamefont {Sharma}},\
  }\bibfield  {title} {\bibinfo {title} {Fraction of delocalized eigenstates in
  the long-range aubry-andr\'e-harper model},\ }\href
  {https://doi.org/10.1103/PhysRevB.103.075124} {\bibfield  {journal} {\bibinfo
   {journal} {Phys. Rev. B}\ }\textbf {\bibinfo {volume} {103}},\ \bibinfo
  {pages} {075124} (\bibinfo {year} {2021})}\BibitemShut {NoStop}%
\bibitem [{\citenamefont {Yao}\ \emph {et~al.}(2019)\citenamefont {Yao},
  \citenamefont {Khoudli}, \citenamefont {Bresque},\ and\ \citenamefont
  {Sanchez-Palencia}}]{PhysRevLett.123.070405}%
  \BibitemOpen
  \bibfield  {author} {\bibinfo {author} {\bibfnamefont {H.}~\bibnamefont
  {Yao}}, \bibinfo {author} {\bibfnamefont {A.}~\bibnamefont {Khoudli}},
  \bibinfo {author} {\bibfnamefont {L.}~\bibnamefont {Bresque}},\ and\ \bibinfo
  {author} {\bibfnamefont {L.}~\bibnamefont {Sanchez-Palencia}},\ }\bibfield
  {title} {\bibinfo {title} {Critical behavior and fractality in shallow
  one-dimensional quasiperiodic potentials},\ }\href
  {https://doi.org/10.1103/PhysRevLett.123.070405} {\bibfield  {journal}
  {\bibinfo  {journal} {Phys. Rev. Lett.}\ }\textbf {\bibinfo {volume} {123}},\
  \bibinfo {pages} {070405} (\bibinfo {year} {2019})}\BibitemShut {NoStop}%
\bibitem [{\citenamefont {Liu}\ \emph {et~al.}(2022)\citenamefont {Liu},
  \citenamefont {Xia}, \citenamefont {Longhi},\ and\ \citenamefont
  {Sanchez-Palencia}}]{10.21468/SciPostPhys.12.1.027}%
  \BibitemOpen
  \bibfield  {author} {\bibinfo {author} {\bibfnamefont {T.}~\bibnamefont
  {Liu}}, \bibinfo {author} {\bibfnamefont {X.}~\bibnamefont {Xia}}, \bibinfo
  {author} {\bibfnamefont {S.}~\bibnamefont {Longhi}},\ and\ \bibinfo {author}
  {\bibfnamefont {L.}~\bibnamefont {Sanchez-Palencia}},\ }\bibfield  {title}
  {\bibinfo {title} {{Anomalous mobility edges in one-dimensional quasiperiodic
  models}},\ }\href {https://doi.org/10.21468/SciPostPhys.12.1.027} {\bibfield
  {journal} {\bibinfo  {journal} {SciPost Phys.}\ }\textbf {\bibinfo {volume}
  {12}},\ \bibinfo {pages} {027} (\bibinfo {year} {2022})}\BibitemShut
  {NoStop}%
\bibitem [{\citenamefont {Wang}\ \emph {et~al.}(2022)\citenamefont {Wang},
  \citenamefont {Zhang}, \citenamefont {Li}, \citenamefont {Wu}, \citenamefont
  {Liu}, \citenamefont {Mei}, \citenamefont {Hu}, \citenamefont {Xiao},
  \citenamefont {Ma}, \citenamefont {Chin},\ and\ \citenamefont
  {Jia}}]{PhysRevLett.129.103401}%
  \BibitemOpen
  \bibfield  {author} {\bibinfo {author} {\bibfnamefont {Y.}~\bibnamefont
  {Wang}}, \bibinfo {author} {\bibfnamefont {J.-H.}\ \bibnamefont {Zhang}},
  \bibinfo {author} {\bibfnamefont {Y.}~\bibnamefont {Li}}, \bibinfo {author}
  {\bibfnamefont {J.}~\bibnamefont {Wu}}, \bibinfo {author} {\bibfnamefont
  {W.}~\bibnamefont {Liu}}, \bibinfo {author} {\bibfnamefont {F.}~\bibnamefont
  {Mei}}, \bibinfo {author} {\bibfnamefont {Y.}~\bibnamefont {Hu}}, \bibinfo
  {author} {\bibfnamefont {L.}~\bibnamefont {Xiao}}, \bibinfo {author}
  {\bibfnamefont {J.}~\bibnamefont {Ma}}, \bibinfo {author} {\bibfnamefont
  {C.}~\bibnamefont {Chin}},\ and\ \bibinfo {author} {\bibfnamefont
  {S.}~\bibnamefont {Jia}},\ }\bibfield  {title} {\bibinfo {title} {Observation
  of interaction-induced mobility edge in an atomic aubry-andr\'e wire},\
  }\href {https://doi.org/10.1103/PhysRevLett.129.103401} {\bibfield  {journal}
  {\bibinfo  {journal} {Phys. Rev. Lett.}\ }\textbf {\bibinfo {volume} {129}},\
  \bibinfo {pages} {103401} (\bibinfo {year} {2022})}\BibitemShut {NoStop}%
\bibitem [{\citenamefont {An}\ \emph {et~al.}(2021)\citenamefont {An},
  \citenamefont {Padavi\ifmmode~\acute{c}\else \'{c}\fi{}}, \citenamefont
  {Meier}, \citenamefont {Hegde}, \citenamefont {Ganeshan}, \citenamefont
  {Pixley}, \citenamefont {Vishveshwara},\ and\ \citenamefont
  {Gadway}}]{PhysRevLett.126.040603}%
  \BibitemOpen
  \bibfield  {author} {\bibinfo {author} {\bibfnamefont {F.~A.}\ \bibnamefont
  {An}}, \bibinfo {author} {\bibfnamefont {K.}~\bibnamefont
  {Padavi\ifmmode~\acute{c}\else \'{c}\fi{}}}, \bibinfo {author} {\bibfnamefont
  {E.~J.}\ \bibnamefont {Meier}}, \bibinfo {author} {\bibfnamefont
  {S.}~\bibnamefont {Hegde}}, \bibinfo {author} {\bibfnamefont
  {S.}~\bibnamefont {Ganeshan}}, \bibinfo {author} {\bibfnamefont {J.~H.}\
  \bibnamefont {Pixley}}, \bibinfo {author} {\bibfnamefont {S.}~\bibnamefont
  {Vishveshwara}},\ and\ \bibinfo {author} {\bibfnamefont {B.}~\bibnamefont
  {Gadway}},\ }\bibfield  {title} {\bibinfo {title} {Interactions and mobility
  edges: Observing the generalized aubry-andr\'e model},\ }\href
  {https://doi.org/10.1103/PhysRevLett.126.040603} {\bibfield  {journal}
  {\bibinfo  {journal} {Phys. Rev. Lett.}\ }\textbf {\bibinfo {volume} {126}},\
  \bibinfo {pages} {040603} (\bibinfo {year} {2021})}\BibitemShut {NoStop}%
\bibitem [{\citenamefont {Ashida}\ \emph {et~al.}(2020)\citenamefont {Ashida},
  \citenamefont {Gong},\ and\ \citenamefont {Ueda}}]{ashida2020non}%
  \BibitemOpen
  \bibfield  {author} {\bibinfo {author} {\bibfnamefont {Y.}~\bibnamefont
  {Ashida}}, \bibinfo {author} {\bibfnamefont {Z.}~\bibnamefont {Gong}},\ and\
  \bibinfo {author} {\bibfnamefont {M.}~\bibnamefont {Ueda}},\ }\bibfield
  {title} {\bibinfo {title} {Non-{H}ermitian physics},\ }\href
  {https://doi.org/10.1080/00018732.2021.1876991} {\bibfield  {journal}
  {\bibinfo  {journal} {Advances in Physics}\ }\textbf {\bibinfo {volume}
  {69}},\ \bibinfo {pages} {249} (\bibinfo {year} {2020})},\ \Eprint
  {https://arxiv.org/abs/https://doi.org/10.1080/00018732.2021.1876991}
  {https://doi.org/10.1080/00018732.2021.1876991} \BibitemShut {NoStop}%
\bibitem [{\citenamefont {Bergholtz}\ \emph {et~al.}(2021)\citenamefont
  {Bergholtz}, \citenamefont {Budich},\ and\ \citenamefont
  {Kunst}}]{bergholtz2021exceptional}%
  \BibitemOpen
  \bibfield  {author} {\bibinfo {author} {\bibfnamefont {E.~J.}\ \bibnamefont
  {Bergholtz}}, \bibinfo {author} {\bibfnamefont {J.~C.}\ \bibnamefont
  {Budich}},\ and\ \bibinfo {author} {\bibfnamefont {F.~K.}\ \bibnamefont
  {Kunst}},\ }\bibfield  {title} {\bibinfo {title} {Exceptional topology of
  non-{H}ermitian systems},\ }\href
  {https://doi.org/10.1103/RevModPhys.93.015005} {\bibfield  {journal}
  {\bibinfo  {journal} {Rev. Mod. Phys.}\ }\textbf {\bibinfo {volume} {93}},\
  \bibinfo {pages} {015005} (\bibinfo {year} {2021})}\BibitemShut {NoStop}%
\bibitem [{\citenamefont {Wang}\ \emph
  {et~al.}(2021{\natexlab{a}})\citenamefont {Wang}, \citenamefont {Zhang},
  \citenamefont {Hua}, \citenamefont {Lei}, \citenamefont {Lu},\ and\
  \citenamefont {Chen}}]{wang2021topological}%
  \BibitemOpen
  \bibfield  {author} {\bibinfo {author} {\bibfnamefont {H.}~\bibnamefont
  {Wang}}, \bibinfo {author} {\bibfnamefont {X.}~\bibnamefont {Zhang}},
  \bibinfo {author} {\bibfnamefont {J.}~\bibnamefont {Hua}}, \bibinfo {author}
  {\bibfnamefont {D.}~\bibnamefont {Lei}}, \bibinfo {author} {\bibfnamefont
  {M.}~\bibnamefont {Lu}},\ and\ \bibinfo {author} {\bibfnamefont
  {Y.}~\bibnamefont {Chen}},\ }\bibfield  {title} {\bibinfo {title}
  {Topological physics of non-{H}ermitian optics and photonics: a review},\
  }\href {https://doi.org/10.1088/2040-8986/ac2e15} {\bibfield  {journal}
  {\bibinfo  {journal} {Journal of Optics}\ }\textbf {\bibinfo {volume} {23}},\
  \bibinfo {pages} {123001} (\bibinfo {year} {2021}{\natexlab{a}})}\BibitemShut
  {NoStop}%
\bibitem [{\citenamefont {Li}\ \emph {et~al.}(2020)\citenamefont {Li},
  \citenamefont {Li}, \citenamefont {Zhang},\ and\ \citenamefont
  {Gong}}]{li2020symmetry}%
  \BibitemOpen
  \bibfield  {author} {\bibinfo {author} {\bibfnamefont {X.-S.}\ \bibnamefont
  {Li}}, \bibinfo {author} {\bibfnamefont {Z.-Z.}\ \bibnamefont {Li}}, \bibinfo
  {author} {\bibfnamefont {L.-L.}\ \bibnamefont {Zhang}},\ and\ \bibinfo
  {author} {\bibfnamefont {W.-J.}\ \bibnamefont {Gong}},\ }\bibfield  {title}
  {\bibinfo {title} {$\mathcal{PT}$-symmetry of the {S}u-{S}chrieffer-{H}eeger
  model with imaginary boundary potentials and next-nearest-neighboring
  coupling},\ }\href {https://doi.org/10.1088/1361-648x/ab62bd} {\bibfield
  {journal} {\bibinfo  {journal} {Journal of Physics: Condensed Matter}\
  }\textbf {\bibinfo {volume} {32}},\ \bibinfo {pages} {165401} (\bibinfo
  {year} {2020})}\BibitemShut {NoStop}%
\bibitem [{\citenamefont {Yuce}\ and\ \citenamefont
  {Oztas}(2018)}]{yuce2018pt}%
  \BibitemOpen
  \bibfield  {author} {\bibinfo {author} {\bibfnamefont {C.}~\bibnamefont
  {Yuce}}\ and\ \bibinfo {author} {\bibfnamefont {Z.}~\bibnamefont {Oztas}},\
  }\bibfield  {title} {\bibinfo {title} {{PT} symmetry protected
  non-{H}ermitian topological systems},\ }\href
  {https://doi.org/10.1038/s41598-018-35795-5} {\bibfield  {journal} {\bibinfo
  {journal} {Scientific reports}\ }\textbf {\bibinfo {volume} {8}},\ \bibinfo
  {pages} {1} (\bibinfo {year} {2018})}\BibitemShut {NoStop}%
\bibitem [{\citenamefont {Jin}\ \emph {et~al.}(2017)\citenamefont {Jin},
  \citenamefont {Wang},\ and\ \citenamefont {Song}}]{jin2017schrieffer}%
  \BibitemOpen
  \bibfield  {author} {\bibinfo {author} {\bibfnamefont {L.}~\bibnamefont
  {Jin}}, \bibinfo {author} {\bibfnamefont {P.}~\bibnamefont {Wang}},\ and\
  \bibinfo {author} {\bibfnamefont {Z.}~\bibnamefont {Song}},\ }\bibfield
  {title} {\bibinfo {title} {Su-{S}chrieffer-{H}eeger chain with one pair of
  {PT}-symmetric defects},\ }\href {https://doi.org/10.1038/s41598-017-06198-9}
  {\bibfield  {journal} {\bibinfo  {journal} {Scientific Reports}\ }\textbf
  {\bibinfo {volume} {7}},\ \bibinfo {pages} {1} (\bibinfo {year}
  {2017})}\BibitemShut {NoStop}%
\bibitem [{\citenamefont {Zhu}\ \emph {et~al.}(2014)\citenamefont {Zhu},
  \citenamefont {L\"u},\ and\ \citenamefont {Chen}}]{zhu2014pt}%
  \BibitemOpen
  \bibfield  {author} {\bibinfo {author} {\bibfnamefont {B.}~\bibnamefont
  {Zhu}}, \bibinfo {author} {\bibfnamefont {R.}~\bibnamefont {L\"u}},\ and\
  \bibinfo {author} {\bibfnamefont {S.}~\bibnamefont {Chen}},\ }\bibfield
  {title} {\bibinfo {title} {$\mathcal{PT}$ symmetry in the non-{H}ermitian
  {S}u-{S}chrieffer-{H}eeger model with complex boundary potentials},\ }\href
  {https://doi.org/10.1103/PhysRevA.89.062102} {\bibfield  {journal} {\bibinfo
  {journal} {Phys. Rev. A}\ }\textbf {\bibinfo {volume} {89}},\ \bibinfo
  {pages} {062102} (\bibinfo {year} {2014})}\BibitemShut {NoStop}%
\bibitem [{\citenamefont {Xu}\ \emph {et~al.}(2020)\citenamefont {Xu},
  \citenamefont {Zhang}, \citenamefont {Chen}, \citenamefont {Fu},\ and\
  \citenamefont {Zhang}}]{xu2020fate}%
  \BibitemOpen
  \bibfield  {author} {\bibinfo {author} {\bibfnamefont {Z.}~\bibnamefont
  {Xu}}, \bibinfo {author} {\bibfnamefont {R.}~\bibnamefont {Zhang}}, \bibinfo
  {author} {\bibfnamefont {S.}~\bibnamefont {Chen}}, \bibinfo {author}
  {\bibfnamefont {L.}~\bibnamefont {Fu}},\ and\ \bibinfo {author}
  {\bibfnamefont {Y.}~\bibnamefont {Zhang}},\ }\bibfield  {title} {\bibinfo
  {title} {Fate of zero modes in a finite {S}u-{S}chrieffer-{H}eeger model with
  $\mathcal{PT}$ symmetry},\ }\href
  {https://doi.org/10.1103/PhysRevA.101.013635} {\bibfield  {journal} {\bibinfo
   {journal} {Phys. Rev. A}\ }\textbf {\bibinfo {volume} {101}},\ \bibinfo
  {pages} {013635} (\bibinfo {year} {2020})}\BibitemShut {NoStop}%
\bibitem [{\citenamefont {Borgnia}\ \emph {et~al.}(2020)\citenamefont
  {Borgnia}, \citenamefont {Kruchkov},\ and\ \citenamefont
  {Slager}}]{PhysRevLett.124.056802}%
  \BibitemOpen
  \bibfield  {author} {\bibinfo {author} {\bibfnamefont {D.~S.}\ \bibnamefont
  {Borgnia}}, \bibinfo {author} {\bibfnamefont {A.~J.}\ \bibnamefont
  {Kruchkov}},\ and\ \bibinfo {author} {\bibfnamefont {R.-J.}\ \bibnamefont
  {Slager}},\ }\bibfield  {title} {\bibinfo {title} {Non-hermitian boundary
  modes and topology},\ }\href {https://doi.org/10.1103/PhysRevLett.124.056802}
  {\bibfield  {journal} {\bibinfo  {journal} {Phys. Rev. Lett.}\ }\textbf
  {\bibinfo {volume} {124}},\ \bibinfo {pages} {056802} (\bibinfo {year}
  {2020})}\BibitemShut {NoStop}%
\bibitem [{\citenamefont {Okuma}\ \emph {et~al.}(2020)\citenamefont {Okuma},
  \citenamefont {Kawabata}, \citenamefont {Shiozaki},\ and\ \citenamefont
  {Sato}}]{PhysRevLett.124.086801}%
  \BibitemOpen
  \bibfield  {author} {\bibinfo {author} {\bibfnamefont {N.}~\bibnamefont
  {Okuma}}, \bibinfo {author} {\bibfnamefont {K.}~\bibnamefont {Kawabata}},
  \bibinfo {author} {\bibfnamefont {K.}~\bibnamefont {Shiozaki}},\ and\
  \bibinfo {author} {\bibfnamefont {M.}~\bibnamefont {Sato}},\ }\bibfield
  {title} {\bibinfo {title} {Topological origin of non-{H}ermitian skin
  effects},\ }\href {https://doi.org/10.1103/PhysRevLett.124.086801} {\bibfield
   {journal} {\bibinfo  {journal} {Phys. Rev. Lett.}\ }\textbf {\bibinfo
  {volume} {124}},\ \bibinfo {pages} {086801} (\bibinfo {year}
  {2020})}\BibitemShut {NoStop}%
\bibitem [{\citenamefont {Zhang}\ \emph {et~al.}(2020)\citenamefont {Zhang},
  \citenamefont {Yang},\ and\ \citenamefont {Fang}}]{PhysRevLett.125.126402}%
  \BibitemOpen
  \bibfield  {author} {\bibinfo {author} {\bibfnamefont {K.}~\bibnamefont
  {Zhang}}, \bibinfo {author} {\bibfnamefont {Z.}~\bibnamefont {Yang}},\ and\
  \bibinfo {author} {\bibfnamefont {C.}~\bibnamefont {Fang}},\ }\bibfield
  {title} {\bibinfo {title} {Correspondence between winding numbers and skin
  modes in non-{H}ermitian systems},\ }\href
  {https://doi.org/10.1103/PhysRevLett.125.126402} {\bibfield  {journal}
  {\bibinfo  {journal} {Phys. Rev. Lett.}\ }\textbf {\bibinfo {volume} {125}},\
  \bibinfo {pages} {126402} (\bibinfo {year} {2020})}\BibitemShut {NoStop}%
\bibitem [{\citenamefont {Yao}\ and\ \citenamefont {Wang}(2018)}]{yao2018edge}%
  \BibitemOpen
  \bibfield  {author} {\bibinfo {author} {\bibfnamefont {S.}~\bibnamefont
  {Yao}}\ and\ \bibinfo {author} {\bibfnamefont {Z.}~\bibnamefont {Wang}},\
  }\bibfield  {title} {\bibinfo {title} {Edge states and topological invariants
  of non-{H}ermitian systems},\ }\href
  {https://doi.org/10.1103/PhysRevLett.121.086803} {\bibfield  {journal}
  {\bibinfo  {journal} {Phys. Rev. Lett.}\ }\textbf {\bibinfo {volume} {121}},\
  \bibinfo {pages} {086803} (\bibinfo {year} {2018})}\BibitemShut {NoStop}%
\bibitem [{\citenamefont {Song}\ \emph {et~al.}(2019)\citenamefont {Song},
  \citenamefont {Yao},\ and\ \citenamefont {Wang}}]{song2019non}%
  \BibitemOpen
  \bibfield  {author} {\bibinfo {author} {\bibfnamefont {F.}~\bibnamefont
  {Song}}, \bibinfo {author} {\bibfnamefont {S.}~\bibnamefont {Yao}},\ and\
  \bibinfo {author} {\bibfnamefont {Z.}~\bibnamefont {Wang}},\ }\bibfield
  {title} {\bibinfo {title} {Non-{H}ermitian topological invariants in real
  space},\ }\href {https://doi.org/10.1103/PhysRevLett.123.246801} {\bibfield
  {journal} {\bibinfo  {journal} {Phys. Rev. Lett.}\ }\textbf {\bibinfo
  {volume} {123}},\ \bibinfo {pages} {246801} (\bibinfo {year}
  {2019})}\BibitemShut {NoStop}%
\bibitem [{\citenamefont {Lee}(2016)}]{lee2016anomalous}%
  \BibitemOpen
  \bibfield  {author} {\bibinfo {author} {\bibfnamefont {T.~E.}\ \bibnamefont
  {Lee}},\ }\bibfield  {title} {\bibinfo {title} {Anomalous edge state in a
  non-{H}ermitian lattice},\ }\href
  {https://doi.org/10.1103/PhysRevLett.116.133903} {\bibfield  {journal}
  {\bibinfo  {journal} {Phys. Rev. Lett.}\ }\textbf {\bibinfo {volume} {116}},\
  \bibinfo {pages} {133903} (\bibinfo {year} {2016})}\BibitemShut {NoStop}%
\bibitem [{\citenamefont {Xia}\ \emph {et~al.}(2022)\citenamefont {Xia},
  \citenamefont {Huang}, \citenamefont {Wang},\ and\ \citenamefont
  {Li}}]{PhysRevB.105.014207}%
  \BibitemOpen
  \bibfield  {author} {\bibinfo {author} {\bibfnamefont {X.}~\bibnamefont
  {Xia}}, \bibinfo {author} {\bibfnamefont {K.}~\bibnamefont {Huang}}, \bibinfo
  {author} {\bibfnamefont {S.}~\bibnamefont {Wang}},\ and\ \bibinfo {author}
  {\bibfnamefont {X.}~\bibnamefont {Li}},\ }\bibfield  {title} {\bibinfo
  {title} {Exact mobility edges in the non-hermitian
  ${t}_{1}\text{\ensuremath{-}}{t}_{2}$ model: Theory and possible experimental
  realizations},\ }\href {https://doi.org/10.1103/PhysRevB.105.014207}
  {\bibfield  {journal} {\bibinfo  {journal} {Phys. Rev. B}\ }\textbf {\bibinfo
  {volume} {105}},\ \bibinfo {pages} {014207} (\bibinfo {year}
  {2022})}\BibitemShut {NoStop}%
\bibitem [{\citenamefont {Liu}\ \emph {et~al.}(2020{\natexlab{a}})\citenamefont
  {Liu}, \citenamefont {Guo}, \citenamefont {Pu},\ and\ \citenamefont
  {Longhi}}]{PhysRevB.102.024205}%
  \BibitemOpen
  \bibfield  {author} {\bibinfo {author} {\bibfnamefont {T.}~\bibnamefont
  {Liu}}, \bibinfo {author} {\bibfnamefont {H.}~\bibnamefont {Guo}}, \bibinfo
  {author} {\bibfnamefont {Y.}~\bibnamefont {Pu}},\ and\ \bibinfo {author}
  {\bibfnamefont {S.}~\bibnamefont {Longhi}},\ }\bibfield  {title} {\bibinfo
  {title} {Generalized aubry-andr\'e self-duality and mobility edges in
  non-hermitian quasiperiodic lattices},\ }\href
  {https://doi.org/10.1103/PhysRevB.102.024205} {\bibfield  {journal} {\bibinfo
   {journal} {Phys. Rev. B}\ }\textbf {\bibinfo {volume} {102}},\ \bibinfo
  {pages} {024205} (\bibinfo {year} {2020}{\natexlab{a}})}\BibitemShut
  {NoStop}%
\bibitem [{\citenamefont {Wang}\ \emph
  {et~al.}(2021{\natexlab{b}})\citenamefont {Wang}, \citenamefont {Xia},
  \citenamefont {Wang}, \citenamefont {Zheng},\ and\ \citenamefont
  {Liu}}]{PhysRevB.103.174205}%
  \BibitemOpen
  \bibfield  {author} {\bibinfo {author} {\bibfnamefont {Y.}~\bibnamefont
  {Wang}}, \bibinfo {author} {\bibfnamefont {X.}~\bibnamefont {Xia}}, \bibinfo
  {author} {\bibfnamefont {Y.}~\bibnamefont {Wang}}, \bibinfo {author}
  {\bibfnamefont {Z.}~\bibnamefont {Zheng}},\ and\ \bibinfo {author}
  {\bibfnamefont {X.-J.}\ \bibnamefont {Liu}},\ }\bibfield  {title} {\bibinfo
  {title} {Duality between two generalized aubry-andr\'e models with exact
  mobility edges},\ }\href {https://doi.org/10.1103/PhysRevB.103.174205}
  {\bibfield  {journal} {\bibinfo  {journal} {Phys. Rev. B}\ }\textbf {\bibinfo
  {volume} {103}},\ \bibinfo {pages} {174205} (\bibinfo {year}
  {2021}{\natexlab{b}})}\BibitemShut {NoStop}%
\bibitem [{\citenamefont {Zhou}\ and\ \citenamefont
  {Gu}(2021)}]{Zhou2021TopologicalDT}%
  \BibitemOpen
  \bibfield  {author} {\bibinfo {author} {\bibfnamefont {L.}~\bibnamefont
  {Zhou}}\ and\ \bibinfo {author} {\bibfnamefont {Y.}~\bibnamefont {Gu}},\
  }\bibfield  {title} {\bibinfo {title} {Topological delocalization transitions
  and mobility edges in the nonreciprocal maryland model},\ }\href
  {https://api.semanticscholar.org/CorpusID:237091598} {\bibfield  {journal}
  {\bibinfo  {journal} {Journal of Physics: Condensed Matter}\ }\textbf
  {\bibinfo {volume} {34}} (\bibinfo {year} {2021})}\BibitemShut {NoStop}%
\bibitem [{\citenamefont {Zhai}\ \emph {et~al.}(2021)\citenamefont {Zhai},
  \citenamefont {Huang},\ and\ \citenamefont {Yin}}]{PhysRevB.104.014202}%
  \BibitemOpen
  \bibfield  {author} {\bibinfo {author} {\bibfnamefont {L.-J.}\ \bibnamefont
  {Zhai}}, \bibinfo {author} {\bibfnamefont {G.-Y.}\ \bibnamefont {Huang}},\
  and\ \bibinfo {author} {\bibfnamefont {S.}~\bibnamefont {Yin}},\ }\bibfield
  {title} {\bibinfo {title} {Cascade of the delocalization transition in a
  non-hermitian interpolating aubry-andr\'e-fibonacci chain},\ }\href
  {https://doi.org/10.1103/PhysRevB.104.014202} {\bibfield  {journal} {\bibinfo
   {journal} {Phys. Rev. B}\ }\textbf {\bibinfo {volume} {104}},\ \bibinfo
  {pages} {014202} (\bibinfo {year} {2021})}\BibitemShut {NoStop}%
\bibitem [{\citenamefont {Liu}\ \emph {et~al.}(2020{\natexlab{b}})\citenamefont
  {Liu}, \citenamefont {Jiang}, \citenamefont {Cao},\ and\ \citenamefont
  {Chen}}]{PhysRevB.101.174205}%
  \BibitemOpen
  \bibfield  {author} {\bibinfo {author} {\bibfnamefont {Y.}~\bibnamefont
  {Liu}}, \bibinfo {author} {\bibfnamefont {X.-P.}\ \bibnamefont {Jiang}},
  \bibinfo {author} {\bibfnamefont {J.}~\bibnamefont {Cao}},\ and\ \bibinfo
  {author} {\bibfnamefont {S.}~\bibnamefont {Chen}},\ }\bibfield  {title}
  {\bibinfo {title} {Non-hermitian mobility edges in one-dimensional
  quasicrystals with parity-time symmetry},\ }\href
  {https://doi.org/10.1103/PhysRevB.101.174205} {\bibfield  {journal} {\bibinfo
   {journal} {Phys. Rev. B}\ }\textbf {\bibinfo {volume} {101}},\ \bibinfo
  {pages} {174205} (\bibinfo {year} {2020}{\natexlab{b}})}\BibitemShut
  {NoStop}%
\bibitem [{\citenamefont {Xu}\ \emph {et~al.}(2021)\citenamefont {Xu},
  \citenamefont {Xia},\ and\ \citenamefont {Chen}}]{PhysRevB.104.224204}%
  \BibitemOpen
  \bibfield  {author} {\bibinfo {author} {\bibfnamefont {Z.}~\bibnamefont
  {Xu}}, \bibinfo {author} {\bibfnamefont {X.}~\bibnamefont {Xia}},\ and\
  \bibinfo {author} {\bibfnamefont {S.}~\bibnamefont {Chen}},\ }\bibfield
  {title} {\bibinfo {title} {Non-hermitian aubry-andr\'e model with power-law
  hopping},\ }\href {https://doi.org/10.1103/PhysRevB.104.224204} {\bibfield
  {journal} {\bibinfo  {journal} {Phys. Rev. B}\ }\textbf {\bibinfo {volume}
  {104}},\ \bibinfo {pages} {224204} (\bibinfo {year} {2021})}\BibitemShut
  {NoStop}%
\bibitem [{\citenamefont {Peng}\ \emph {et~al.}(2023)\citenamefont {Peng},
  \citenamefont {Cheng},\ and\ \citenamefont {Xianlong}}]{PhysRevB.107.174205}%
  \BibitemOpen
  \bibfield  {author} {\bibinfo {author} {\bibfnamefont {D.}~\bibnamefont
  {Peng}}, \bibinfo {author} {\bibfnamefont {S.}~\bibnamefont {Cheng}},\ and\
  \bibinfo {author} {\bibfnamefont {G.}~\bibnamefont {Xianlong}},\ }\bibfield
  {title} {\bibinfo {title} {Power law hopping of single particles in
  one-dimensional non-hermitian quasicrystals},\ }\href
  {https://doi.org/10.1103/PhysRevB.107.174205} {\bibfield  {journal} {\bibinfo
   {journal} {Phys. Rev. B}\ }\textbf {\bibinfo {volume} {107}},\ \bibinfo
  {pages} {174205} (\bibinfo {year} {2023})}\BibitemShut {NoStop}%
\bibitem [{\citenamefont {Cai}\ \emph {et~al.}(2013)\citenamefont {Cai},
  \citenamefont {Lang}, \citenamefont {Chen},\ and\ \citenamefont
  {Wang}}]{PhysRevLett.110.176403}%
  \BibitemOpen
  \bibfield  {author} {\bibinfo {author} {\bibfnamefont {X.}~\bibnamefont
  {Cai}}, \bibinfo {author} {\bibfnamefont {L.-J.}\ \bibnamefont {Lang}},
  \bibinfo {author} {\bibfnamefont {S.}~\bibnamefont {Chen}},\ and\ \bibinfo
  {author} {\bibfnamefont {Y.}~\bibnamefont {Wang}},\ }\bibfield  {title}
  {\bibinfo {title} {Topological superconductor to anderson localization
  transition in one-dimensional incommensurate lattices},\ }\href
  {https://doi.org/10.1103/PhysRevLett.110.176403} {\bibfield  {journal}
  {\bibinfo  {journal} {Phys. Rev. Lett.}\ }\textbf {\bibinfo {volume} {110}},\
  \bibinfo {pages} {176403} (\bibinfo {year} {2013})}\BibitemShut {NoStop}%
\bibitem [{\citenamefont {DeGottardi}\ \emph {et~al.}(2013)\citenamefont
  {DeGottardi}, \citenamefont {Sen},\ and\ \citenamefont
  {Vishveshwara}}]{PhysRevLett.110.146404}%
  \BibitemOpen
  \bibfield  {author} {\bibinfo {author} {\bibfnamefont {W.}~\bibnamefont
  {DeGottardi}}, \bibinfo {author} {\bibfnamefont {D.}~\bibnamefont {Sen}},\
  and\ \bibinfo {author} {\bibfnamefont {S.}~\bibnamefont {Vishveshwara}},\
  }\bibfield  {title} {\bibinfo {title} {Majorana fermions in superconducting
  1d systems having periodic, quasiperiodic, and disordered potentials},\
  }\href {https://doi.org/10.1103/PhysRevLett.110.146404} {\bibfield  {journal}
  {\bibinfo  {journal} {Phys. Rev. Lett.}\ }\textbf {\bibinfo {volume} {110}},\
  \bibinfo {pages} {146404} (\bibinfo {year} {2013})}\BibitemShut {NoStop}%
\bibitem [{\citenamefont {Fraxanet}\ \emph {et~al.}(2022)\citenamefont
  {Fraxanet}, \citenamefont {Bhattacharya}, \citenamefont {Grass},
  \citenamefont {Lewenstein},\ and\ \citenamefont
  {Dauphin}}]{PhysRevB.106.024204}%
  \BibitemOpen
  \bibfield  {author} {\bibinfo {author} {\bibfnamefont {J.}~\bibnamefont
  {Fraxanet}}, \bibinfo {author} {\bibfnamefont {U.}~\bibnamefont
  {Bhattacharya}}, \bibinfo {author} {\bibfnamefont {T.}~\bibnamefont {Grass}},
  \bibinfo {author} {\bibfnamefont {M.}~\bibnamefont {Lewenstein}},\ and\
  \bibinfo {author} {\bibfnamefont {A.}~\bibnamefont {Dauphin}},\ }\bibfield
  {title} {\bibinfo {title} {Localization and multifractal properties of the
  long-range kitaev chain in the presence of an aubry-andr\'e-harper
  modulation},\ }\href {https://doi.org/10.1103/PhysRevB.106.024204} {\bibfield
   {journal} {\bibinfo  {journal} {Phys. Rev. B}\ }\textbf {\bibinfo {volume}
  {106}},\ \bibinfo {pages} {024204} (\bibinfo {year} {2022})}\BibitemShut
  {NoStop}%
\bibitem [{\citenamefont {Zeng}\ \emph {et~al.}(2018)\citenamefont {Zeng},
  \citenamefont {Chen},\ and\ \citenamefont {Lü}}]{Zeng_2018}%
  \BibitemOpen
  \bibfield  {author} {\bibinfo {author} {\bibfnamefont {Q.-B.}\ \bibnamefont
  {Zeng}}, \bibinfo {author} {\bibfnamefont {S.}~\bibnamefont {Chen}},\ and\
  \bibinfo {author} {\bibfnamefont {R.}~\bibnamefont {L\"u}},\ }\bibfield
  {title} {\bibinfo {title} {Quench dynamics in the aubry-andr\'e-{H}arper
  model with p-wave superconductivity},\ }\href
  {https://doi.org/10.1088/1367-2630/aabe39} {\bibfield  {journal} {\bibinfo
  {journal} {New Journal of Physics}\ }\textbf {\bibinfo {volume} {20}},\
  \bibinfo {pages} {053012} (\bibinfo {year} {2018})}\BibitemShut {NoStop}%
\bibitem [{\citenamefont {Yahyavi}\ \emph {et~al.}(2019)\citenamefont
  {Yahyavi}, \citenamefont {Het\'enyi},\ and\ \citenamefont
  {Tanatar}}]{PhysRevB.100.064202}%
  \BibitemOpen
  \bibfield  {author} {\bibinfo {author} {\bibfnamefont {M.}~\bibnamefont
  {Yahyavi}}, \bibinfo {author} {\bibfnamefont {B.}~\bibnamefont {Het\'enyi}},\
  and\ \bibinfo {author} {\bibfnamefont {B.}~\bibnamefont {Tanatar}},\
  }\bibfield  {title} {\bibinfo {title} {Generalized aubry-andr\'e-harper model
  with modulated hopping and $p$-wave pairing},\ }\href
  {https://doi.org/10.1103/PhysRevB.100.064202} {\bibfield  {journal} {\bibinfo
   {journal} {Phys. Rev. B}\ }\textbf {\bibinfo {volume} {100}},\ \bibinfo
  {pages} {064202} (\bibinfo {year} {2019})}\BibitemShut {NoStop}%
\bibitem [{\citenamefont {Tong}\ \emph {et~al.}(2021)\citenamefont {Tong},
  \citenamefont {Meng}, \citenamefont {Jiang}, \citenamefont {Lee},
  \citenamefont {Neto},\ and\ \citenamefont {Xianlong}}]{PhysRevB.103.104202}%
  \BibitemOpen
  \bibfield  {author} {\bibinfo {author} {\bibfnamefont {X.}~\bibnamefont
  {Tong}}, \bibinfo {author} {\bibfnamefont {Y.-M.}\ \bibnamefont {Meng}},
  \bibinfo {author} {\bibfnamefont {X.}~\bibnamefont {Jiang}}, \bibinfo
  {author} {\bibfnamefont {C.}~\bibnamefont {Lee}}, \bibinfo {author}
  {\bibfnamefont {G.~D. d.~M.}\ \bibnamefont {Neto}},\ and\ \bibinfo {author}
  {\bibfnamefont {G.}~\bibnamefont {Xianlong}},\ }\bibfield  {title} {\bibinfo
  {title} {Dynamics of a quantum phase transition in the aubry-andr\'e-harper
  model with $p$-wave superconductivity},\ }\href
  {https://doi.org/10.1103/PhysRevB.103.104202} {\bibfield  {journal} {\bibinfo
   {journal} {Phys. Rev. B}\ }\textbf {\bibinfo {volume} {103}},\ \bibinfo
  {pages} {104202} (\bibinfo {year} {2021})}\BibitemShut {NoStop}%
\bibitem [{\citenamefont {Lv}\ \emph {et~al.}(2022)\citenamefont {Lv},
  \citenamefont {Yi}, \citenamefont {Li}, \citenamefont {Sun},\ and\
  \citenamefont {You}}]{PhysRevA.105.013315}%
  \BibitemOpen
  \bibfield  {author} {\bibinfo {author} {\bibfnamefont {T.}~\bibnamefont
  {Lv}}, \bibinfo {author} {\bibfnamefont {T.-C.}\ \bibnamefont {Yi}}, \bibinfo
  {author} {\bibfnamefont {L.}~\bibnamefont {Li}}, \bibinfo {author}
  {\bibfnamefont {G.}~\bibnamefont {Sun}},\ and\ \bibinfo {author}
  {\bibfnamefont {W.-L.}\ \bibnamefont {You}},\ }\bibfield  {title} {\bibinfo
  {title} {Quantum criticality and universality in the $p$-wave-paired
  aubry-andr\'e-harper model},\ }\href
  {https://doi.org/10.1103/PhysRevA.105.013315} {\bibfield  {journal} {\bibinfo
   {journal} {Phys. Rev. A}\ }\textbf {\bibinfo {volume} {105}},\ \bibinfo
  {pages} {013315} (\bibinfo {year} {2022})}\BibitemShut {NoStop}%
\bibitem [{\citenamefont {Wang}\ \emph {et~al.}(2016)\citenamefont {Wang},
  \citenamefont {Liu}, \citenamefont {Xianlong},\ and\ \citenamefont
  {Hu}}]{PhysRevB.93.104504}%
  \BibitemOpen
  \bibfield  {author} {\bibinfo {author} {\bibfnamefont {J.}~\bibnamefont
  {Wang}}, \bibinfo {author} {\bibfnamefont {X.-J.}\ \bibnamefont {Liu}},
  \bibinfo {author} {\bibfnamefont {G.}~\bibnamefont {Xianlong}},\ and\
  \bibinfo {author} {\bibfnamefont {H.}~\bibnamefont {Hu}},\ }\bibfield
  {title} {\bibinfo {title} {Phase diagram of a non-abelian
  aubry-andr\'e-harper model with $p$-wave superfluidity},\ }\href
  {https://doi.org/10.1103/PhysRevB.93.104504} {\bibfield  {journal} {\bibinfo
  {journal} {Phys. Rev. B}\ }\textbf {\bibinfo {volume} {93}},\ \bibinfo
  {pages} {104504} (\bibinfo {year} {2016})}\BibitemShut {NoStop}%
\bibitem [{\citenamefont {Zeng}\ \emph {et~al.}(2017)\citenamefont {Zeng},
  \citenamefont {Chen},\ and\ \citenamefont {L\"u}}]{PhysRevA.95.062118}%
  \BibitemOpen
  \bibfield  {author} {\bibinfo {author} {\bibfnamefont {Q.-B.}\ \bibnamefont
  {Zeng}}, \bibinfo {author} {\bibfnamefont {S.}~\bibnamefont {Chen}},\ and\
  \bibinfo {author} {\bibfnamefont {R.}~\bibnamefont {L\"u}},\ }\bibfield
  {title} {\bibinfo {title} {Anderson localization in the non-hermitian
  aubry-andr\'e-harper model with physical gain and loss},\ }\href
  {https://doi.org/10.1103/PhysRevA.95.062118} {\bibfield  {journal} {\bibinfo
  {journal} {Phys. Rev. A}\ }\textbf {\bibinfo {volume} {95}},\ \bibinfo
  {pages} {062118} (\bibinfo {year} {2017})}\BibitemShut {NoStop}%
\bibitem [{\citenamefont {Cai}(2021)}]{PhysRevB.103.214202}%
  \BibitemOpen
  \bibfield  {author} {\bibinfo {author} {\bibfnamefont {X.}~\bibnamefont
  {Cai}},\ }\bibfield  {title} {\bibinfo {title} {Localization and topological
  phase transitions in non-hermitian aubry-andr\'e-harper models with $p$-wave
  pairing},\ }\href {https://doi.org/10.1103/PhysRevB.103.214202} {\bibfield
  {journal} {\bibinfo  {journal} {Phys. Rev. B}\ }\textbf {\bibinfo {volume}
  {103}},\ \bibinfo {pages} {214202} (\bibinfo {year} {2021})}\BibitemShut
  {NoStop}%
\bibitem [{\citenamefont {Gandhi}\ and\ \citenamefont
  {Bandyopadhyay}(2023)}]{PhysRevB.108.014204}%
  \BibitemOpen
  \bibfield  {author} {\bibinfo {author} {\bibfnamefont {S.}~\bibnamefont
  {Gandhi}}\ and\ \bibinfo {author} {\bibfnamefont {J.~N.}\ \bibnamefont
  {Bandyopadhyay}},\ }\bibfield  {title} {\bibinfo {title} {Topological triple
  phase transition in non-hermitian quasicrystals with complex asymmetric
  hopping},\ }\href {https://doi.org/10.1103/PhysRevB.108.014204} {\bibfield
  {journal} {\bibinfo  {journal} {Phys. Rev. B}\ }\textbf {\bibinfo {volume}
  {108}},\ \bibinfo {pages} {014204} (\bibinfo {year} {2023})}\BibitemShut
  {NoStop}%
\bibitem [{\citenamefont {Liu}\ \emph {et~al.}(2021{\natexlab{a}})\citenamefont
  {Liu}, \citenamefont {Cheng}, \citenamefont {Guo},\ and\ \citenamefont
  {Xianlong}}]{PhysRevB.103.104203}%
  \BibitemOpen
  \bibfield  {author} {\bibinfo {author} {\bibfnamefont {T.}~\bibnamefont
  {Liu}}, \bibinfo {author} {\bibfnamefont {S.}~\bibnamefont {Cheng}}, \bibinfo
  {author} {\bibfnamefont {H.}~\bibnamefont {Guo}},\ and\ \bibinfo {author}
  {\bibfnamefont {G.}~\bibnamefont {Xianlong}},\ }\bibfield  {title} {\bibinfo
  {title} {Fate of majorana zero modes, exact location of critical states, and
  unconventional real-complex transition in non-hermitian quasiperiodic
  lattices},\ }\href {https://doi.org/10.1103/PhysRevB.103.104203} {\bibfield
  {journal} {\bibinfo  {journal} {Phys. Rev. B}\ }\textbf {\bibinfo {volume}
  {103}},\ \bibinfo {pages} {104203} (\bibinfo {year}
  {2021}{\natexlab{a}})}\BibitemShut {NoStop}%
\bibitem [{\citenamefont {Longhi}(2019)}]{PhysRevLett.122.237601}%
  \BibitemOpen
  \bibfield  {author} {\bibinfo {author} {\bibfnamefont {S.}~\bibnamefont
  {Longhi}},\ }\bibfield  {title} {\bibinfo {title} {Topological phase
  transition in non-hermitian quasicrystals},\ }\href
  {https://doi.org/10.1103/PhysRevLett.122.237601} {\bibfield  {journal}
  {\bibinfo  {journal} {Phys. Rev. Lett.}\ }\textbf {\bibinfo {volume} {122}},\
  \bibinfo {pages} {237601} (\bibinfo {year} {2019})}\BibitemShut {NoStop}%
\bibitem [{\citenamefont {Longhi}(2023)}]{PhysRevB.108.075121}%
  \BibitemOpen
  \bibfield  {author} {\bibinfo {author} {\bibfnamefont {S.}~\bibnamefont
  {Longhi}},\ }\bibfield  {title} {\bibinfo {title} {Phase transitions and
  bunching of correlated particles in a non-hermitian quasicrystal},\ }\href
  {https://doi.org/10.1103/PhysRevB.108.075121} {\bibfield  {journal} {\bibinfo
   {journal} {Phys. Rev. B}\ }\textbf {\bibinfo {volume} {108}},\ \bibinfo
  {pages} {075121} (\bibinfo {year} {2023})}\BibitemShut {NoStop}%
\bibitem [{\citenamefont {Liu}\ \emph {et~al.}(2021{\natexlab{b}})\citenamefont
  {Liu}, \citenamefont {Zhou},\ and\ \citenamefont
  {Chen}}]{PhysRevB.104.024201}%
  \BibitemOpen
  \bibfield  {author} {\bibinfo {author} {\bibfnamefont {Y.}~\bibnamefont
  {Liu}}, \bibinfo {author} {\bibfnamefont {Q.}~\bibnamefont {Zhou}},\ and\
  \bibinfo {author} {\bibfnamefont {S.}~\bibnamefont {Chen}},\ }\bibfield
  {title} {\bibinfo {title} {Localization transition, spectrum structure, and
  winding numbers for one-dimensional non-hermitian quasicrystals},\ }\href
  {https://doi.org/10.1103/PhysRevB.104.024201} {\bibfield  {journal} {\bibinfo
   {journal} {Phys. Rev. B}\ }\textbf {\bibinfo {volume} {104}},\ \bibinfo
  {pages} {024201} (\bibinfo {year} {2021}{\natexlab{b}})}\BibitemShut
  {NoStop}%
\bibitem [{\citenamefont {Bender}\ and\ \citenamefont
  {Boettcher}(1998)}]{bender1998real}%
  \BibitemOpen
  \bibfield  {author} {\bibinfo {author} {\bibfnamefont {C.~M.}\ \bibnamefont
  {Bender}}\ and\ \bibinfo {author} {\bibfnamefont {S.}~\bibnamefont
  {Boettcher}},\ }\bibfield  {title} {\bibinfo {title} {Real spectra in
  non-{H}ermitian {H}amiltonians having $\mathcal{P}\mathcal{T}$ symmetry},\
  }\href {https://doi.org/10.1103/PhysRevLett.80.5243} {\bibfield  {journal}
  {\bibinfo  {journal} {Phys. Rev. Lett.}\ }\textbf {\bibinfo {volume} {80}},\
  \bibinfo {pages} {5243} (\bibinfo {year} {1998})}\BibitemShut {NoStop}%
\bibitem [{\citenamefont {Fleckenstein}\ \emph {et~al.}(2018)\citenamefont
  {Fleckenstein}, \citenamefont {Dom\'{\i}nguez}, \citenamefont
  {Traverso~Ziani},\ and\ \citenamefont {Trauzettel}}]{PhysRevB.97.155425}%
  \BibitemOpen
  \bibfield  {author} {\bibinfo {author} {\bibfnamefont {C.}~\bibnamefont
  {Fleckenstein}}, \bibinfo {author} {\bibfnamefont {F.}~\bibnamefont
  {Dom\'{\i}nguez}}, \bibinfo {author} {\bibfnamefont {N.}~\bibnamefont
  {Traverso~Ziani}},\ and\ \bibinfo {author} {\bibfnamefont {B.}~\bibnamefont
  {Trauzettel}},\ }\bibfield  {title} {\bibinfo {title} {Decaying spectral
  oscillations in a majorana wire with finite coherence length},\ }\href
  {https://doi.org/10.1103/PhysRevB.97.155425} {\bibfield  {journal} {\bibinfo
  {journal} {Phys. Rev. B}\ }\textbf {\bibinfo {volume} {97}},\ \bibinfo
  {pages} {155425} (\bibinfo {year} {2018})}\BibitemShut {NoStop}%
\bibitem [{\citenamefont {Stanescu}\ \emph {et~al.}(2013)\citenamefont
  {Stanescu}, \citenamefont {Lutchyn},\ and\ \citenamefont
  {Das~Sarma}}]{PhysRevB.87.094518}%
  \BibitemOpen
  \bibfield  {author} {\bibinfo {author} {\bibfnamefont {T.~D.}\ \bibnamefont
  {Stanescu}}, \bibinfo {author} {\bibfnamefont {R.~M.}\ \bibnamefont
  {Lutchyn}},\ and\ \bibinfo {author} {\bibfnamefont {S.}~\bibnamefont
  {Das~Sarma}},\ }\bibfield  {title} {\bibinfo {title} {Dimensional crossover
  in spin-orbit-coupled semiconductor nanowires with induced superconducting
  pairing},\ }\href {https://doi.org/10.1103/PhysRevB.87.094518} {\bibfield
  {journal} {\bibinfo  {journal} {Phys. Rev. B}\ }\textbf {\bibinfo {volume}
  {87}},\ \bibinfo {pages} {094518} (\bibinfo {year} {2013})}\BibitemShut
  {NoStop}%
\bibitem [{\citenamefont {Zhang}\ \emph {et~al.}(2022)\citenamefont {Zhang},
  \citenamefont {Nie},\ and\ \citenamefont {Liu}}]{PhysRevApplied.18.024038}%
  \BibitemOpen
  \bibfield  {author} {\bibinfo {author} {\bibfnamefont {Y.}~\bibnamefont
  {Zhang}}, \bibinfo {author} {\bibfnamefont {W.}~\bibnamefont {Nie}},\ and\
  \bibinfo {author} {\bibfnamefont {Y.-x.}\ \bibnamefont {Liu}},\ }\bibfield
  {title} {\bibinfo {title} {Edge-state oscillations in a one-dimensional
  topological chain with dissipative couplings},\ }\href
  {https://doi.org/10.1103/PhysRevApplied.18.024038} {\bibfield  {journal}
  {\bibinfo  {journal} {Phys. Rev. Appl.}\ }\textbf {\bibinfo {volume} {18}},\
  \bibinfo {pages} {024038} (\bibinfo {year} {2022})}\BibitemShut {NoStop}%
\bibitem [{\citenamefont {Vodola}\ \emph {et~al.}(2014)\citenamefont {Vodola},
  \citenamefont {Lepori}, \citenamefont {Ercolessi}, \citenamefont {Gorshkov},\
  and\ \citenamefont {Pupillo}}]{PhysRevLett.113.156402}%
  \BibitemOpen
  \bibfield  {author} {\bibinfo {author} {\bibfnamefont {D.}~\bibnamefont
  {Vodola}}, \bibinfo {author} {\bibfnamefont {L.}~\bibnamefont {Lepori}},
  \bibinfo {author} {\bibfnamefont {E.}~\bibnamefont {Ercolessi}}, \bibinfo
  {author} {\bibfnamefont {A.~V.}\ \bibnamefont {Gorshkov}},\ and\ \bibinfo
  {author} {\bibfnamefont {G.}~\bibnamefont {Pupillo}},\ }\bibfield  {title}
  {\bibinfo {title} {Kitaev chains with long-range pairing},\ }\href
  {https://doi.org/10.1103/PhysRevLett.113.156402} {\bibfield  {journal}
  {\bibinfo  {journal} {Phys. Rev. Lett.}\ }\textbf {\bibinfo {volume} {113}},\
  \bibinfo {pages} {156402} (\bibinfo {year} {2014})}\BibitemShut {NoStop}%
\bibitem [{\citenamefont {Vodola}\ \emph {et~al.}(2015)\citenamefont {Vodola},
  \citenamefont {Lepori}, \citenamefont {Ercolessi},\ and\ \citenamefont
  {Pupillo}}]{Vodola_2016}%
  \BibitemOpen
  \bibfield  {author} {\bibinfo {author} {\bibfnamefont {D.}~\bibnamefont
  {Vodola}}, \bibinfo {author} {\bibfnamefont {L.}~\bibnamefont {Lepori}},
  \bibinfo {author} {\bibfnamefont {E.}~\bibnamefont {Ercolessi}},\ and\
  \bibinfo {author} {\bibfnamefont {G.}~\bibnamefont {Pupillo}},\ }\bibfield
  {title} {\bibinfo {title} {Long-range ising and kitaev models: phases,
  correlations and edge modes},\ }\href
  {https://doi.org/10.1088/1367-2630/18/1/015001} {\bibfield  {journal}
  {\bibinfo  {journal} {New Journal of Physics}\ }\textbf {\bibinfo {volume}
  {18}},\ \bibinfo {pages} {015001} (\bibinfo {year} {2015})}\BibitemShut
  {NoStop}%
\bibitem [{\citenamefont {Viyuela}\ \emph {et~al.}(2016)\citenamefont
  {Viyuela}, \citenamefont {Vodola}, \citenamefont {Pupillo},\ and\
  \citenamefont {Martin-Delgado}}]{PhysRevB.94.125121}%
  \BibitemOpen
  \bibfield  {author} {\bibinfo {author} {\bibfnamefont {O.}~\bibnamefont
  {Viyuela}}, \bibinfo {author} {\bibfnamefont {D.}~\bibnamefont {Vodola}},
  \bibinfo {author} {\bibfnamefont {G.}~\bibnamefont {Pupillo}},\ and\ \bibinfo
  {author} {\bibfnamefont {M.~A.}\ \bibnamefont {Martin-Delgado}},\ }\bibfield
  {title} {\bibinfo {title} {Topological massive dirac edge modes and
  long-range superconducting hamiltonians},\ }\href
  {https://doi.org/10.1103/PhysRevB.94.125121} {\bibfield  {journal} {\bibinfo
  {journal} {Phys. Rev. B}\ }\textbf {\bibinfo {volume} {94}},\ \bibinfo
  {pages} {125121} (\bibinfo {year} {2016})}\BibitemShut {NoStop}%
\bibitem [{\citenamefont {Fraxanet}\ \emph {et~al.}(2021)\citenamefont
  {Fraxanet}, \citenamefont {Bhattacharya}, \citenamefont {Grass},
  \citenamefont {Rakshit}, \citenamefont {Lewenstein},\ and\ \citenamefont
  {Dauphin}}]{PhysRevResearch.3.013148}%
  \BibitemOpen
  \bibfield  {author} {\bibinfo {author} {\bibfnamefont {J.}~\bibnamefont
  {Fraxanet}}, \bibinfo {author} {\bibfnamefont {U.}~\bibnamefont
  {Bhattacharya}}, \bibinfo {author} {\bibfnamefont {T.}~\bibnamefont {Grass}},
  \bibinfo {author} {\bibfnamefont {D.}~\bibnamefont {Rakshit}}, \bibinfo
  {author} {\bibfnamefont {M.}~\bibnamefont {Lewenstein}},\ and\ \bibinfo
  {author} {\bibfnamefont {A.}~\bibnamefont {Dauphin}},\ }\bibfield  {title}
  {\bibinfo {title} {Topological properties of the long-range kitaev chain with
  aubry-andr\'e-harper modulation},\ }\href
  {https://doi.org/10.1103/PhysRevResearch.3.013148} {\bibfield  {journal}
  {\bibinfo  {journal} {Phys. Rev. Res.}\ }\textbf {\bibinfo {volume} {3}},\
  \bibinfo {pages} {013148} (\bibinfo {year} {2021})}\BibitemShut {NoStop}%
\bibitem [{\citenamefont {Schreiber}\ and\ \citenamefont
  {Grussbach}(1991)}]{PhysRevLett.67.607}%
  \BibitemOpen
  \bibfield  {author} {\bibinfo {author} {\bibfnamefont {M.}~\bibnamefont
  {Schreiber}}\ and\ \bibinfo {author} {\bibfnamefont {H.}~\bibnamefont
  {Grussbach}},\ }\bibfield  {title} {\bibinfo {title} {Multifractal wave
  functions at the anderson transition},\ }\href
  {https://doi.org/10.1103/PhysRevLett.67.607} {\bibfield  {journal} {\bibinfo
  {journal} {Phys. Rev. Lett.}\ }\textbf {\bibinfo {volume} {67}},\ \bibinfo
  {pages} {607} (\bibinfo {year} {1991})}\BibitemShut {NoStop}%
\bibitem [{\citenamefont {Gandhi}\ and\ \citenamefont
  {Bandyopadhyay}(2024)}]{PhysRevB.110.094203}%
  \BibitemOpen
  \bibfield  {author} {\bibinfo {author} {\bibfnamefont {S.}~\bibnamefont
  {Gandhi}}\ and\ \bibinfo {author} {\bibfnamefont {J.~N.}\ \bibnamefont
  {Bandyopadhyay}},\ }\bibfield  {title} {\bibinfo {title} {Non-hermitian
  aubry-andr\'e-harper model with short- and long-range $p$-wave pairing},\
  }\href {https://doi.org/10.1103/PhysRevB.110.094203} {\bibfield  {journal}
  {\bibinfo  {journal} {Phys. Rev. B}\ }\textbf {\bibinfo {volume} {110}},\
  \bibinfo {pages} {094203} (\bibinfo {year} {2024})}\BibitemShut {NoStop}%
\bibitem [{\citenamefont {Hegde}\ and\ \citenamefont
  {Vishveshwara}(2016)}]{PhysRevB.94.115166}%
  \BibitemOpen
  \bibfield  {author} {\bibinfo {author} {\bibfnamefont {S.~S.}\ \bibnamefont
  {Hegde}}\ and\ \bibinfo {author} {\bibfnamefont {S.}~\bibnamefont
  {Vishveshwara}},\ }\bibfield  {title} {\bibinfo {title} {Majorana
  wave-function oscillations, fermion parity switches, and disorder in kitaev
  chains},\ }\href {https://doi.org/10.1103/PhysRevB.94.115166} {\bibfield
  {journal} {\bibinfo  {journal} {Phys. Rev. B}\ }\textbf {\bibinfo {volume}
  {94}},\ \bibinfo {pages} {115166} (\bibinfo {year} {2016})}\BibitemShut
  {NoStop}%
\bibitem [{\citenamefont {Wimmer}(2012)}]{Wimmer}%
  \BibitemOpen
  \bibfield  {author} {\bibinfo {author} {\bibfnamefont {M.}~\bibnamefont
  {Wimmer}},\ }\bibfield  {title} {\bibinfo {title} {Algorithm 923: Efficient
  numerical computation of the pfaffian for dense and banded skew-symmetric
  matrices},\ }\bibfield  {journal} {\bibinfo  {journal} {ACM Trans. Math.
  Softw.}\ }\textbf {\bibinfo {volume} {38}},\ \href
  {https://doi.org/10.1145/2331130.2331138} {10.1145/2331130.2331138} (\bibinfo
  {year} {2012})\BibitemShut {NoStop}%
\end{thebibliography}

%

\end{document}